\documentclass[fleqn,usenatbib]{mnras}

\usepackage{newtxtext,newtxmath}

\usepackage[T1]{fontenc}
\usepackage{amsmath}
\DeclareRobustCommand{\VAN}[3]{#2}
\let\VANthebibliography\thebibliography
\def\thebibliography{\DeclareRobustCommand{\VAN}[3]{##3}\VANthebibliography}

\usepackage{graphicx}	
\usepackage{amsmath}	
\usepackage[normalem]{ulem}
\usepackage{comment}

\def\promedio#1{\langle{#1}\rangle}

\newcommand{\Pcl}{P_{\mathrm{cl}}}   
\newcommand{\Pism}{P_{\mathrm{ISM}}}    
    
\newcommand{\Qext}{Q_{\mathrm{ext}}}    

\newcommand{\Sigmagas}{\Sigma_{\mathrm{gas}}}    
\newcommand{\Sigmatot}{\Sigma_{\mathrm{tot}}}    
\defcitealias{Franco_Cox86}{FC86}

\def\apj{ApJ}
\def\apjl{ApJL}
\def\aap{A\& A}
\def\nat{Nature}
\def\araa{ARA\&A}
\def\mnras{MNRAS}

\def\aj{{AJ}}
\def\apjs{ApJS}
\def\bain{{Bull. Astron. Inst. Netherlands}}
\def\mnras{{MNRAS}}

\def\pasp{{PASP}}

\def\promedio#1{\langle{#1}\rangle}

\title[Molecular Clouds and Star Formation in MW]{Pressure Regulated Formation of Molecular Clouds and Stars: The case of the Milky Way}

\author[Franco et al.]{
José Franco,$^{1}$\thanks{E-mail: pepe@astro.unam.mx}, Aldo Rodríguez-Puebla,$^{1}$ Javier Ballesteros-Paredes$^{2}$, and Manuel Zamora-Avilez$^{3}$
\\
$^{1}$Universidad Nacional Autónoma de México, Instituto de Astronomía, A. P. 70-264, 04510, Ciudad de México, México\\
$^{2}$Universidad Nacional Autónoma de México, Instituto de Radioastronomía y Astrofísica, Antigua Carretera a Pátzcuaro 8701, Ex-Hda, San José de la Huerta\\ 58089 Morelia, Michoacán, México\\
$^{3}$Instituto Nacional de Astrofísica, Óptica y Electrónica, Luis E. Erro 1, 72840 Tonantzintla, Puebla, México\\
}

\date{Accepted XXX. Received YYY; in original form ZZZ}

\pubyear{\the\year{}}

\begin{document}
\label{firstpage}
\pagerange{\pageref{firstpage}--\pageref{lastpage}}
\maketitle

\begin{abstract}

We present a steady-state analytical model for pressure-regulated formation of molecular clouds (MC) and stars (SF) in gaseous galactic disks and apply it to the Milky Way (MW). MC formation depends on midplane interstellar pressure $\Pism$ and metallicity $Z$, and for galactocentric distances $R\gtrsim5$ kpc, $\Pism(R)$ scales approximately linearly with molecular gas surface density $\Sigma_{\rm mol}(R)$. The molecularization of the cold neutral medium (CNM) is due to the opacity of small dust grains that protect the center of the cloud from dissociating radiation when the column density is $\Sigma_d\geq 5\ (Z_\odot/Z)M_\odot\text{ pc}^{-2}$. The H$_2$ formation rate per hydrogen atom is $F\sim10^{-15}(\Pism/P_\odot)T_{100}^{-1/2}\text{s}^{-1}$, and the corresponding formation rate per unit area is $\dot{\Sigma}^{+}_{\rm mol}\sim 5\times10^{-2}\left(\Pism/{P_\odot}\right)T_{100}^{-1/2}M_\odot~\text{kpc}^{-2}~\text{yr}^{-1}$, where $P_\odot$ is the pressure at the solar circle and $T_{100}=T/100\text{ K}$ is the temperature of the cloud. In equilibrium, this equals the molecular gas destruction rate $\dot{\Sigma}^{-}_{\rm mol}$ due to SF. Self-gravity sets in when the column density of a cloud reaches $\Sigma_{\rm sg}=\Sigma_{\rm sg,\odot}(\Pism/P_\odot)^{1/2}$, with $\Sigma_{\rm sg,\odot}\sim30\ M_\odot\ \text{pc}^{-2}$. Given the distribution of $\Pism(R)$ and $Z(R)$ in the MW, the SF process at $5\lesssim R\lesssim11$ kpc follows a two-step track: first, MCs form from CNM gas and then they form stars when self-gravity sets in. The resulting SFR surface density is $\Sigma_\text{SFR}(R)\approx (1.6-4)\times10^{-3}\left(\Pism/P_\odot\right)\ \text{M}_\odot~\text {kpc}^{-2}\text{yr}^{-1}$ with an average final SF efficiency of $\epsilon_{\rm sf}\sim (3-8)\times 10^{-2}$.

\end{abstract}







\section{Introduction}

Star formation remains an unsolved problem in astrophysics; although we have a general framework for how stars form, many of the specific mechanisms and outcomes remain poorly understood \cite[see reviews][]{Kennicutt_Evans12, Sanchez_2020, Schinnerer_Leroy2024}. Key open questions include issues from what determines the formation rate of star-forming clouds to how the mass of these clouds is ultimately distributed among stars. Thus, one of the central unresolved questions is how the physical processes act in concert with the system parameters to regulate the molecular cloud and star formation rates, along with the resulting star formation efficiency.

Almost seven decades ago, \citet{Schmidt59} proposed that the rate of star formation should scale as a power of the gas density. Although several studies in the 1970s and early 1980s attempted to quantify this relationship \citep[i. e.,][]{Sanduleak69, Madore+74, Talbot80}, it was not until the late 1980s and 1990s that a clear power-law correlation was established between the surface density of star formation and the surface density of interstellar gas \citep[][ see also \citealp{Kennicutt_DeLosReyes21} for recent determinations]{ Buat+89, Kennicutt89, Kennicutt98}.

Numerous studies in the early 2000s further investigated this correlation, which is usually referred to as the Kennicutt-Schmidt relation, consistently finding similar results but with different gas tracers. These efforts included analyses of the radial dependence of the star formation rate on the total gas surface density in galaxies \citep[e.g.,][]{Wong_Blitz2002, Blitz_Rosolowsky04, Blitz_Rosolowsky_2006, Sanchez_2020}, measurements of the star formation rate as a function of the mass in dense gas in unresolved galaxies \citep[e.g.,][]{Gao-Solomon04}, and studies of galaxies resolved at kiloparsec scales \citet{Bigiel+2008}. Interestingly, a variety of correlations have been found between the measurements of the star formation rate with molecular gas mass and interstellar pressure \citep[e.g., ][For comprehensive reviews of these works, see, e.g., \citealt{Kennicutt_Evans12, Sanchez_2020, Schinnerer_Leroy2024}]{Elmegreen1993b, Elmegreen_Parravano1994, Wong_Blitz2002, Blitz_Rosolowsky04, Blitz_Rosolowsky_2006,Fisher+17,Sanchez_2020,Barrera-Ballesteros+21, Sun+23, Ellison+24}.

Various theoretical models have been proposed to explain the observed correlations; for example, \citet{Elmegreen02} sought to explain the Kennicutt-Schmidt relation by assuming that the galactic disks maintain an approximately constant scale height and that the gas is undergoing gravitational collapse. Other studies have interpreted the correlation in terms of large-scale gravitational instabilities, particularly with the \citet{Toomre64} criterion \citep[e.g.,][]{Li+05}, with turbulent support of the interstellar medium \citep[e.g.,][]{Krumholz_McKee05, Padoan+12, Kim+21}, or with sustained gas accretion into molecular clouds \citep{Zamora-Aviles_Vazquez-Semadeni14}. Several authors have also emphasized the critical role of self-gravity in regulating star formation \citep[e.g.,][]{Krumholz+12, Elmegreen18}, and the role of interstellar pressure in regulating the properties of the interstellar gas and hence star formation \citep[e.g.,][]{Elmegreen1993, Elmegreen_Parravano1994, Wong_Blitz2002, Blitz_Rosolowsky04,Blitz_Rosolowsky_2006,Ostriker+10}. More recently, \citet{Ballesteros-Paredes+24} proposed that the diversity of observational correlations reported in the literature may arise naturally from the gravitational collapse of dense gas, with the variations resulting from the specific biases inherent in each observational methodology.

An idea that has received more attention is the self-consistent model proposed by \citet{Ostriker+10}, which has been further developed in a series of articles that include numerical simulations \citep[see][]{Kim+2011, Kim+2017, Ostriker_Kim22}. The model assumes that star formation is self-regulated, where the main source of energy for the interstellar medium comes from recently formed stars, keeping the ionization and temperature of the gas components in balance, along with their velocity dispersions and corresponding scale heights. The resulting star formation rate then compensates for the energy and momentum losses of the system, and the gaseous disk is maintained in a stable dynamical equilibrium. With these assumptions, the star formation rate per unit area is the one required to provide support against the weight of a column density of gas, and, as a result, the rate becomes proportional to the disk pressure at midplane. For pressures below log($\Pism/k)\sim 5.5$, this result is in accordance with the observationally derived linear correlation between star formation rates and the pressures obtained in several surveys of nearby galaxies \citep{Leroy+08,sun+20,Barrera-Ballesteros+21, Ellison+24}. For higher pressures, the correlation is no longer linear, becomes shallower, and models with variations that include the possibility of radial gas inflows as an additional source of kinetic energy for the gas have been proposed by \citet{Krumholz+2018}. 

One issue that remains unanswered in these and other models is how the formation rates of molecular gas and stars are entangled and regulated by the overall system properties and how they are distributed along the galactic disk. Here we address these and other related questions with a simplified steady-state analytical model for a gaseous galactic disk and apply it to the Milky Way (MW). We derive the dependence of molecular cloud formation and star formation as a function of two basic parameters of the system, the interstellar pressure in the midplane, $\Pism$, and its metallicity, $Z$. In Section \ref{sec:pressure_and_clouds} we use the available data to obtain the interstellar pressure in the midplane as a function of the galactocentric radius. Then, assuming that the population of cold diffuse clouds are the precursors of molecular clouds, in Section \ref{sec:molecular_clouds} we derive the radial variation of the threshold conditions for the formation of molecular clouds, the formation rate of H$_2$ on dust grains as a function of $\Pism$, and the formation rate of molecular gas per unit surface along the disk. In Section \ref{sec:selg_grav_clouds_and_sfr},  the threshold conditions for self-gravity are presented, and we obtain the star formation rate per unit surface along with its corresponding star formation efficiency. The model is simple and basic but provides a broad picture of the role that pressure and metallicity play in the regulation of the formation process of both molecular gas and stars.


\section{System pressure and diffuse clouds in the Milky Way}
\label{sec:pressure_and_clouds}

\subsection{Pressure and density of diffuse clouds}

The interstellar pressure in the disk of a spiral galaxy is due to the weight of the column density of its interstellar gas. For a multi-component disk (with gaseous, stellar, and dark-matter components), locally plane-parallel, in rotational equilibrium and in steady state, Poisson's equation in cylindrical coordinates, 
\begin{equation}
    \frac{1}{r} \frac{\partial}{\partial r} \left( r \frac{\partial \Phi}{\partial r} \right) + \frac{1}{r^2} \frac{\partial^2 \Phi}{\partial \theta^2} + \frac{\partial^2 \Phi}{\partial z^2} = -4\pi G\rho_{\rm tot},
    \label{eq:poisson_cylindrical}
\end{equation} 
has a simple solution for the pressure of a thin gaseous disk at any given galatocentric radius: the gravitational acceleration due to the gravitational potential $\Phi$ depends only on $z$, is perpendicular to the disk, and the interstellar pressure is given by
\begin{equation}
    \frac{d\Pism}{dz}=-\rho_g g_z,
\end{equation}
where $G$ is the gravitational constant, $\rho_{\rm tot}$ is the total mass volume density, $\rho_g$ is the gas mass volume density, and $g_z$ is the local gravitational acceleration along the $z$ axis. The pressure at midplane along the disk is then
\begin{equation}
\begin{aligned}
    P_\text{ISM} = & \frac{\pi G}{2} \Sigma_\text{gas}\Sigma_\text{tot} = \frac{\pi G}{2}\Sigma_\text{gas}\left[\Sigma_\text{gas} + (\Sigma_*+\Sigma_\text{dm})\right]\\
    &
= \frac{\pi G}{2}\left(\Sigma_\text{gas}^2 + \Omega \right),
    \end{aligned}
    \label{eq:PISM_eq}
\end{equation}
where $\Sigma_\text{gas}$ is the surface mass density of the three gaseous components (molecular, atomic, and ionized), $\Sigma_\text{tot}$ is the total surface mass density, including the contribution of gas, stars ($\Sigma_*$) and dark matter ($\Sigma_\text{dm}$), and $\Omega=\Sigma_\text{gas}(\Sigma_\ast+\Sigma_\text{dm})$. This steady-state pressure controls the average properties of the ambient gas, and any impulsive transient event (such as the injection of energy from new stellar clusters, the passage of spiral arms, the action of torques driving radial gas inflows, or collisions with high-velocity clouds) can increase the local ISM pressure momentarily, but the system relaxes back to its base value after a crossing time of the perturbed region.

The first term on the right-hand side of Eq.~(\ref{eq:PISM_eq}) is proportional to $\Sigma^2_\text{gas}$ and represents the pressure component due to the gas's own self-gravity. The second term, proportional to $\Omega$, is the dominant component along spiral disks and is due to the gravitational field provided by stars and dark matter. At the solar circle, it is 7 times higher than that of the gas self-gravity (see below), and it has been approximated in two different ways to account for the difference in scale heights of gas and stars in external galaxies (but neglecting the contribution from dark matter). In one case, assuming hydrostatic equilibrium, the ratio of gas and star velocity dispersions, $\sigma_\text{gas}/\sigma_{\ast,z}$, has been used to approximate this difference \citep{Elmegreen1989,Elmegreen_Parravano1994,Wong_Blitz2002}, 
\begin{equation}
    \Omega_H \approx \Sigma_\text{gas} \frac{\sigma_\text{gas}}{\sigma_{\ast,z}}\Sigma_\ast.
\end{equation}
The resulting total pressure is usually termed the hydrostatic equilibrium pressure $P_{\rm H}$. In the second case, the energy input from star formation is used to maintain the velocity dispersion and the scale height of the ambient gas. The stellar mass is assumed to be distributed in an isothermal disk (with average volume density $\rho_*$), where the stellar scale height $h_*$ and the stellar velocity dispersion are related by $\sigma_{*,z}\approx (2\pi  G\Sigma_* h_*)^{1/2}$. These assumptions give \citep[see][]{Blitz_Rosolowsky04,Ostriker+10,Kim+2011,sun+20}, 
\begin{equation}
    \Omega_{DE} \approx \frac{2}{\pi G}\Sigma_\text{gas} \ \sigma_{\text{gas},z}\sqrt{2G\rho_\ast},
\end{equation}
and the corresponding pressure is called the dynamical equilibrium pressure $P_{\rm DE}$. Note that these definitions differ only by a constant, $\Omega_{DE}/\Omega_H=(16/\pi)^{1/2}$, and the resulting values of the dynamical equilibrium pressure are approximately two times higher than those of the hydrostatic equilibrium \citep[see Table 2 in][]{Fisher+2019}. Our definition of interstellar pressure includes the contribution of dark matter and, except for a factor very close to unity (1.05), is equal to the one termed gravitational pressure $P_{\rm grav}$ by \citet{Wilson+2019} and \citet{Molina+2020}.

Regardless of these definitions, the weight of the gas is largely supported by the volume-averaged thermal pressure of the gaseous disk and the nonthermal pressures ascribed to Galactic cosmic rays, turbulent velocities and magnetic fields, all of which seem to have similar energy densities in the solar neighborhood and are generally assumed to be in equipartition along the disk \citep[see][]{Boulares_Cox1990,Ferriere2001,Cox2005,Grenier+2015,Commercon+2019,Crocker+2021}. However, a key question for understanding the possible role of nonthermal pressures in the formation of molecular clouds and stars is how their ambient energy densities are coupled and transmitted to existing diffuse clouds, whose internal pressures must be in balance with the external medium. This is a relevant issue because the thermal pressures of diffuse clouds, which are their best known physical properties, are clearly below the external pressures \citep[see][]{Jenkins_Tripp2001}. 

Cosmic rays and magnetic fields are closely coupled, but the details of the collective trapping and coupling of cosmic rays in clouds depend on the diffusion coefficient along the magnetic field lines (which in turn depends on the cosmic ray energies and the strength of the turbulent component of the magnetic field). Individually, they provide ionization and heating and drive the chemistry in dark clouds, but their collective effects can influence the structure and evolution not only of the cloud population but also of the diffuse ionized medium and may even launch galactic winds \citep[see e.g.,][]{Grenier+2015,Girichidis+2016,Commercon+2019,Armillotta+2021,Hopkins+2020,Brugaletta+2025,RodriguezMontero+2024,Barreto-Mota+2024}. In particular, numerical simulations of cosmic ray feedback in turbulent and magnetized interstellar media indicate that, when the diffusion coefficient parallel to the magnetic field is below a critical value $D_{\rm crit}\sim 10^{24}-10^{25} \text{cm}^2 \ \text{s}^{-1}$, they can pair with the evolution of diffuse clouds, providing effective internal support and preventing them from reaching denser states \citep{Commercon+2019}. As the value of the diffusion coefficient increases, the coupling is less and less efficient, letting clouds reach denser and cooler stages. For Galactic cosmic rays at around 1 GeV (the peak of their energy spectrum; \citealp{Grenier+2015}) the diffusion coefficient is about $10^{28}\text{cm}^2 \ \text{s}^{-1}$ \citep{Ptuskin+2006}, well above the critical value, and in circum-galactic media it can reach values of $\sim 3\times10^{29}\text{cm}^2 \ \text{s}^{-1}$ \citep{Hopkins+2020}. A recent study with ``mirror diffusion'' (cosmic ray scattering at large pitch angles that can produce a secondary acceleration in turbulent molecular clouds; see \citealp{Lazarian+2011}) shows that the corresponding diffusion coefficient tends to values around $10^{27}\text{cm}^2 \ \text{s}^{-1}$ \citep{Barreto-Mota+2024}, also above the critical value. These results, then, imply that the coupling with the cold neutral cloud population of the MW disk is weak, which is in line with the fact that our Galaxy displays a bistable diffuse neutral medium \citep[see][]{Wolfire+2003,Commercon+2019}. Here we assume that cosmic rays do not collectively couple with the cold neutral medium (CNM) of the MW and also that the total internal pressures of cold diffuse clouds are in equilibrium with their external ambient pressure $\Pcl=\Pism$.

Magnetic fields and turbulent flows are important in structuring the interstellar medium and reducing the efficiency of star formation, as clearly shown in a series of works with detailed numerical simulations displaying their effects when acting in concert with self-gravity and stellar feedback \citep[e.g., ][]{Federrath_Klessen12, Federrath_Klessen13, Federrath13, Federrath2016}. A recent study by \citet{Seta_McClure-Griffiths25} shows that the observed average values of the magnetic and turbulent energy densities along the line of sight of different sources in our Galaxy, $E_{\rm mag}$ and $E_{\rm kin}$, follow a power law relationship $E_{\rm mag}\propto {E_{\rm kin}}^{\alpha_m} $, where the exponent $\alpha_m$ varies between 0.9 and 0.64. This implies that magnetic fields do not simply scale with gas density, as is usually assumed, but also with the local velocity field, making it difficult to know the actual coupling of these energy densities with the cold neutral phase. This complication also makes it difficult to assess their possible role in the formation of dense molecular clouds. However, as previously discussed, the thermal energy densities of diffuse clouds are better known than any other properties of clouds, and their values are generally a factor of 4 to 8 below the total ambient pressure \citep[see e.g.,][]{Boulares_Cox1990,Jenkins_Tripp2001, Wolfire+2003,Cox2005,Jenkins2009}. Thus, as a reasonable approximation, we set the total CNM cloud pressure in terms of its thermal pressure and assume that
$P_\text{cl} \sim 6 P_\text{th} = 6n_\text{cl} kT_\text{eff}$, where $k$ is the Boltzmann constant and $T_\text{eff}$ is the average effective cloud temperature. With this assumption, the contribution of nonthermal pressures is implicitly included, and the average cloud volume densities can be approximated as
\begin{equation}
    n_\text{cl} \approx \frac{\Pism}{6\ kT_\text{eff}}.
    \label{eq:cloud_density}
\end{equation}

\subsection{Radial distributions in the Milky Way}
\label{sec:MW_radial_distr}

The MW has a boxy peanut-shaped bulge and a stellar bar whose major axis is estimated to be at an angle with the line of sight between $23^\circ$ and $45^\circ$ \citep[see ][and references therein]{Dwek+1995,Mosenkov+2021}. The determinations of the radius of the bar major axis range from 1 to 4 kpc \citep[see][and references therein]{Freudenreich+1998,Zoccali_Valenti2024}, whose effects are clearly present up to 4 kpc. The location of the solar circle is estimated to be within 7.5 and 8.4 kpc of the Galactic center \citep[e. g.,][]{Zhu_Shen2013,Francis_Anderson2014,Reid+2019}, and for simplicity we set $R_{\odot}=8$ kpc. 

Here we will exclude the bulge zone and restrict discussions of the radial distributions outward of 4 kpc, where the rotation curve becomes approximately flat \citep{McGaugh2016}, and the gaseous disk can be considered thin, locally plane-parallel, and in rotational equilibrium. With regard to the dark matter component, which is supposed to be distributed in a spherical halo \citep{NFW}, it may introduce a gravitational radial component along the disk, but we will assume that it is small compared to the total $z$ component of the gravitational acceleration. The data compilations for the solar neighborhood indicate its presence in the disk \citep[e.g.,][see \citealp{Cheng+2024} for a recent discussion of this issue]{Bienayme+2014,McKee+2015,Kramer_Randall2016a,Kramer_Randall2016b}, and its contribution has been included in the determinations of the total mass surface densities $\Sigma_\text{tot}$ (gas, stars, and dark matter).

The derivations of the total surface gas mass densities in the solar circle $\Sigma_\text {gas}(R_{\odot})$, range from 7 to 13 $M_\odot$ pc$^{-2}$, while the total mass surface density integrated up to approximately 1~kpc above the midplane is about $\Sigma_\text{tot} (R_{\odot},|z|\lesssim1 \; \text{kpc})\approx 68 \ M_\odot$ pc$^{-2}$ \citep[see][]{Romano+2000,Bovy_Rix2013,Flynn+2006,Kalberla_Dedes2008,Pineda+2013,Bienayme+2014,McKee+2015,Kramer_Randall2016b, Marasco+2017, Miville-Deschenes+2017,Cheng+2024}. For the surface density of the different gas components, we use 1.5 $M_\odot$ pc$^{-2}$ for the molecular gas \citep{Miville-Deschenes+2017}, 5 $M_\odot$ pc$^{-2}$ for the atomic gas and 2 $M_\odot$ pc$^{-2}$ for the Reynolds ionized layer \citep{Kramer_Randall2016b}. This gives a conservative value of $\Sigma_\text{gas}~(R_{\odot})\approx~8.5$~$M_\odot$~pc$^{-2}$. Our choice for the surface density of the total mass is $\Sigma_\text{tot} (R_{\odot})\approx68~M_\odot$~pc$^{-2}$ because its integration coincides with the scale height of the Reynolds layer \citep[e.g.,][]{Reynolds+1977,Reynolds1993}, which is approximately 1~kpc and, for simplicity, we omit the specification that the integration of the total mass is up to $|z| \sim 1$ kpc. With these values, the total interstellar pressure in the solar neighborhood is approximately $P_\odot~\approx~3~\times~10^{-12}$~dyn~cm$^{-2}$, or $P_\odot/k\approx~2~\times~10^{4}$~cm$^{-3}$ K, which is in the range of values inferred for the total pressure at the solar circle, $(2-4)~\times~10^{-12}$~dyn~cm$^{-2}$ \citep{Boulares_Cox1990,Ferriere2001,Cox2005}.

\begin{figure*}
    \centering
    \includegraphics[width=\columnwidth]{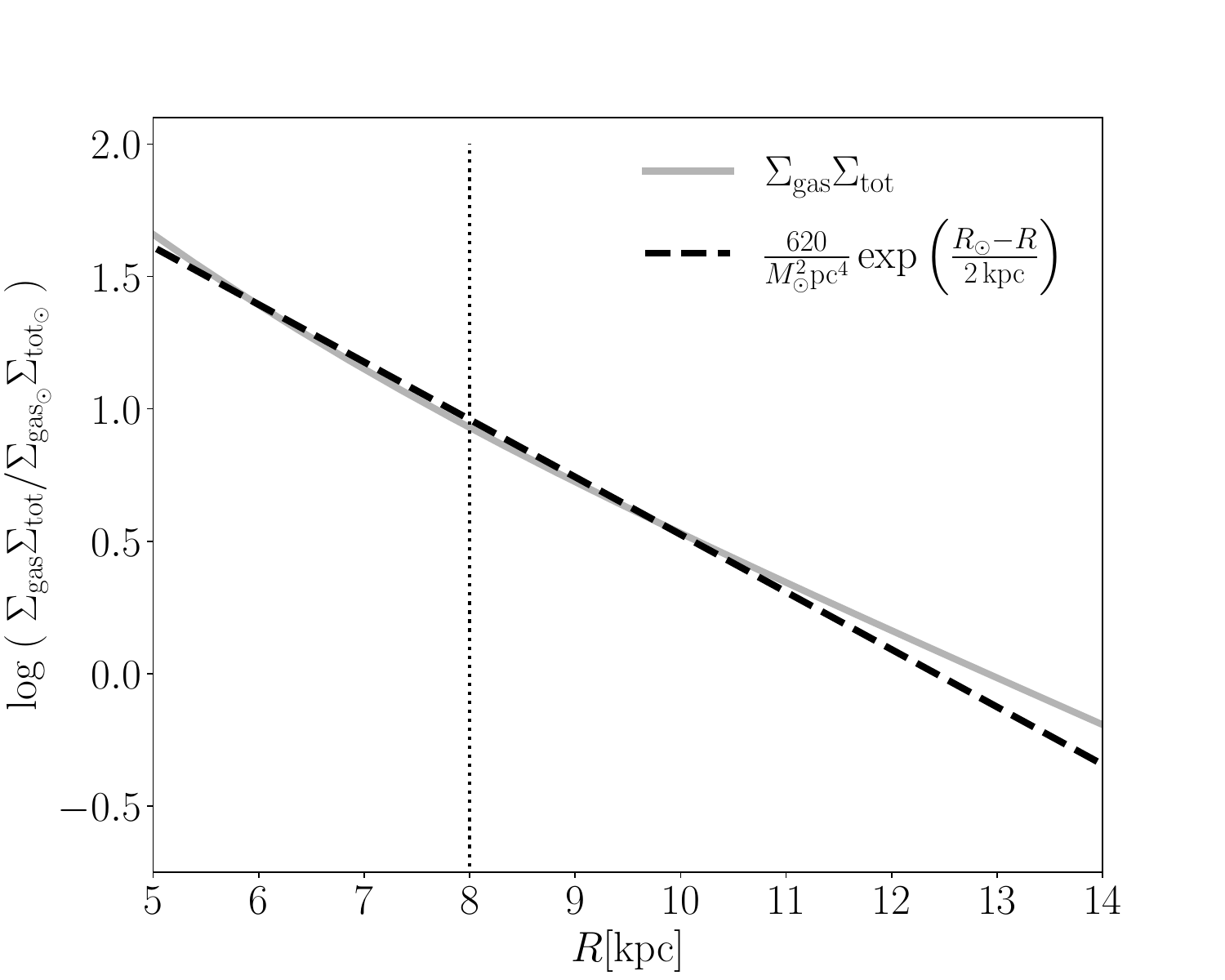}
    \includegraphics[width=\columnwidth]{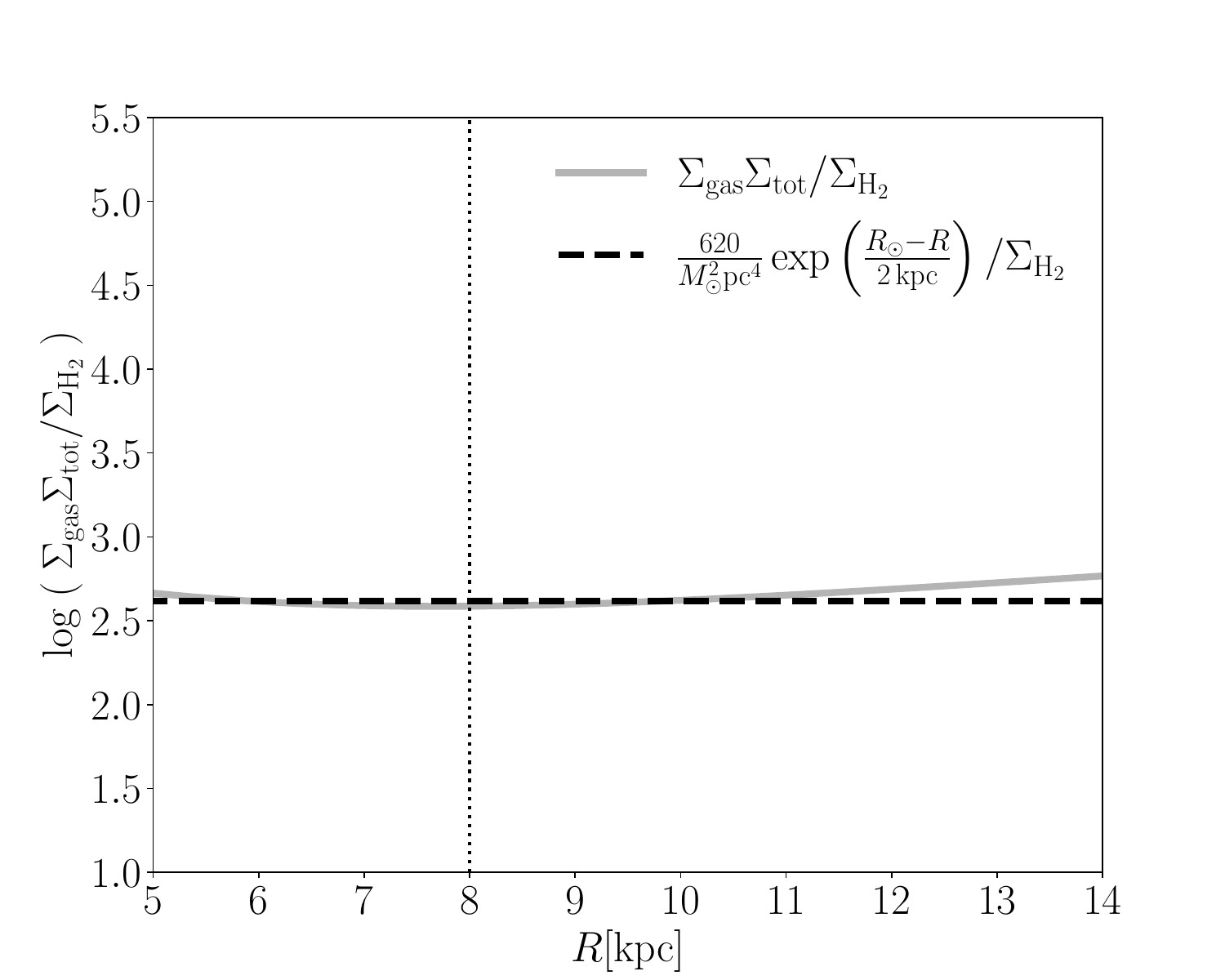}
    \caption{ \textbf{Left Panel:} The product of surface gas mass density and total mass surface density (solid line) and the approximate fit (dashed line), as a function of galactocentric radius. \textbf{Right Panel:} The product of the surface gas mass density and total mass surface density divided by the molecular surface gas mass density. This is proportional to the pressure-H$_2$ relation described in the text and shows that $P_\text{ISM}/\Sigma_{\text{H}_{2}}$ is almost constant between 5 and 12 kpc. In both panels, the dotted lines shows the location of our assumed solar circle, $R_\odot=8$ kpc.}
    \label{fig:Surface_mass_densities}
\end{figure*}

With respect to radial distributions, \citet{Bovy_Rix2013} analyzed the total surface mass density for galactocentric radii from 4.5 to 9~kpc. Their modeled MW disk mass, within 1.1~kpc from the midplane, follows an exponential distribution with a scale length of 2.5~kpc and, as mentioned above, the total surface mass density at the solar circle is 68 $M_\odot$ pc$^{-2}$. The contribution of the stellar component is $\sim 40\ M_\odot$ pc$^{-2}$, resulting in a stellar to total mass ratio of $\sim0.6$. These values are similar to those found by \citet{Bienayme+2014}, and we assume that the radial distribution of $\Sigma_\text{tot}$ can be extrapolated to a galactocentric radius of $\sim$14 kpc, 
\begin{equation}
 \Sigma_\text{tot}(R) \approx 68 \  \text{exp} \left(\frac{R_{\odot}-R}{2.5\ \text{kpc}}\right) M_\odot~\text{pc}^{-2}.
\end{equation} 

The surface gas density in the MW has a peak at the Galactic center and a second peak at the molecular ring at about 4 kpc, then decreases monotonically afterwards, with several peaks associated with the presence of spiral arms \citep[][]{Kalberla_Dedes2008,Kramer_Randall2016b,Pineda+2013,Marasco+2017,Miville-Deschenes+2017}. With respect to each specific gas component, the distribution of the surface mass densities of H$_2$ derived by \citet{Roman-Duval+10} covers the inner MW, while \citet{Kramer_Randall2016b} covers the Galactic radii interval from 4 to 14 kpc and the compilation by \citet{Miville-Deschenes+2017} extends it to almost $R=$20 kpc. The calibration and slopes of the radial distribution outside 5 kpc display several differences \citep[see,][] {Spilker+21}, but for simplicity we use the more extended and explicitly stated exponential fit with the radial scale length of 2 kpc of \citet{Miville-Deschenes+2017}, whose value at the location of the Sun is $\Sigma_{H_2} (R_{\odot})\approx 1.5 \ \ M_\odot \ \text{pc}^{-2}$. Thus, we approximate the surface density of the molecular gas mass outside 4 kpc with 
\begin{equation}
 \Sigma_{H_2}(R) \approx 1.5 \ \text{exp} \left(\frac{R_{\odot}-R}{2 \ \text{kpc}}\right) M_\odot \ \text{pc}^{-2}.
\end{equation}

Regarding the surface mass density of the atomic gas, it has several components with a warm component reaching heights of 400 pc above the midplane \citep[see][]{Dickey_Lockman1990,Ferriere2001,Cox2005}, and its radial distribution is almost flat up to 12 kpc, with most reported values ranging from 4 to 6 $M_\odot$ pc$^{-2}$ (note that \citealp{Kalberla_Dedes2008} report a value of 10 $M_\odot$ pc$^{-2}$, but this is well above the rest of the estimates; see the discussions by \citealp{Kalberla_Dedes2008} and \citealp{Marasco+2017}). Outside 13 kpc, it displays an exponential decrease with a radial scale length of 3.5 kpc. So, we use a constant value of about $\Sigma_{HI}=5\ M_\odot$ pc$^{-2}$ up to 13 kpc, and then assume the outward exponential decrease described by \citet{Kalberla_Dedes2008}. Regarding the diffuse and ionized Reynolds layer \citep[e.g.,][]{Reynolds+1977,Reynolds1993}, it extends up to about 1 kpc from the midplane and appears to have a flat distribution along the disk with a value of 2 $M_\odot$ pc$^{-2}$ \citep[see][]{Kramer_Randall2016b}. Its presence is prevalent as an important extraplanar component in many other galaxies, sometimes extending several kpc above the main disk \citep[e.g.,][]{Dettmar+1990,Collins+2000,Vale_Asari2021}. Adding the column density of the three gas components, we model the total gas distribution at galactocentric radii between 5 and 13 kpc as
\begin{equation}
 \Sigma_\text{gas}(R) \approx 1.5 \ \text{exp} \left(\frac{R_{\odot}-R}{2 \ \text{kpc}}\right) + 7 \ M_\odot \ \text{pc}^{-2}.
\end{equation}
Then, the product of the gas and the total surface mass densities can be approximated by (see Figure \ref{fig:Surface_mass_densities})
\begin{equation}
    \begin{aligned}
        \Sigma_\text{gas} \Sigma_\text{tot} = &
        \left[1.5 \ \text{exp} \left(\frac{R_{\odot}-R}{2 \  \text{kpc}}\right) + 7 \right] 
        \left[68 \ \text{exp}\left(\frac{R_{\odot}-R}{2.5 \  \text{kpc}}\right)\right]  M_\odot^2 \ \text{pc}^{-4} \\
        & \approx {620} \ \text{exp}\left(\frac{R_{\odot}-R}{2 \  \text{kpc}}\right){ M_\odot^2 \  \text{pc}^{-4}} ,
\label{eq:approx_S_gas_S_tot}
 \end{aligned}
\end{equation}
resulting in a radial distribution of the midplane ISM pressure with the same radial scale length as that of the molecular gas,
 \begin{equation}
    \Pism (R)=\frac{\pi G}{2} \Sigma_\text{gas}\Sigma_\text{tot}\approx  P_{\odot}\ \text{exp} \left(\frac{R_{\odot}-R}{2 \  \text{kpc}}\right) ,
 \label{eq:pressure_profile}
 \end{equation} 
therefore, leading to a pressure-H$_2$ relation in MW within 5 and 13 kpc,
 \begin{equation}
 \frac{\Pism}{ \Sigma_{H_2}}\sim \text{constant}\approx \frac{P_{\odot}}{\Sigma_{H_2} (R_{\odot})}.
 \label{eq:pressure_to_h2_is_cte}
 \end{equation} 
Since $\Sigma_{HI}$ is about constant, the ratio $\Sigma_{H_2}/\Sigma_{HI}$ is also proportional to $ P_\text{ISM}$, and is equivalent to the almost linear relationship between the molecular-to-atomic-mass ratio and the system pressure $\Sigma_{H_2}/\Sigma_{HI}\propto P_\text{ISM}^{0.92}$, found by \citet{Blitz_Rosolowsky_2006} for the MW and nearby galaxies.

Another interesting implication of Eq. (\ref{eq:pressure_to_h2_is_cte}) arises when using $\Sigma_{H_2}(R_\odot) \approx 1.5 \times 10^6 \text{ M}_\odot \text{ kpc}^{-2}$, and the depletion time of the well-defined Kennicutt-Schmidt relation for H$_2$ found by \citet{Bigiel+2008}, $\Sigma_\text{SFR}/\Sigma_{H_2}=(7\pm3)\times 10^{-10} \text{yr}^{-1}$: 
\begin{equation}
    \begin{aligned}
        \Sigma_\text{SFR}(R) \sim & \ 10^{-9}\ \frac{P_\text{ISM}}{P_\odot}\Sigma_{H_2}(R_\odot) \text{ M}_\odot \text{ yr}^{-1} \text{ kpc}^{-2}\\\approx & \ 1.5\times10^{-3}\left(\frac{P_\text{ISM}}{P_\odot}\right) \text{ M}_\odot \text{ yr}^{-1} \text{ kpc}^{-2}, 
     \end{aligned}
     \label{eq:pressure_reg_sfr}
\end{equation}
which is also consistent with the slope near unity ($\alpha=0.92$) obtained by \citet{Blitz_Rosolowsky_2006}\footnote{ The value at the solar circle, $1.6\times10^{-3}\text{ M}_\odot \text{ yr}^{-1} \text{ kpc}^{-2}$, is similar to that derived by \citet{Lee_Miville_Murray16}, and a factor of $\sim2$ smaller compared to the normalization derived by \citet{Sun+23}. This difference is within the dispersion derived in this relation of $\sigma=0.33$ dex \citep{Sun+23}, and gives values within those derived from observations; see Section \ref{sec:radia_SFR_distr}}. This relationship represents a lower limit to the rate that is derived from the basic principles and local molecular cloud data discussed in Section \ref{sec:radia_SFR_distr}.

In addition, note that the radial distributions of $\Sigmatot$ and $\Sigmagas$  become about equal at a Galactic radius close to 13 kpc, where the midplane pressure drops to about a 10th of its solar circle value. This location is almost coincidental with the truncation radius of the old stellar disk, 12 kpc \citep{Freudenreich+1998}, and the location where the atomic gas density begins to decrease exponentially at 13 kpc \citep{Kalberla_Dedes2008}. Also, the vertical scale height of the gaseous disk flares up close to this distance, as shown in the compilation of \citet{Bacchini+2019}, and one would expect that the vertical distribution of the gas may change from an exponential function into an isothermal self-gravitating gaseous disk with a sech$^2$ distribution somewhere close to this location \citep[e.g.,][]{Spitzer1942}.

Finally, with eqs. (\ref{eq:cloud_density}) and (\ref{eq:pressure_profile}), the resulting average volume density of diffuse pressure-bounded clouds in the disk can be written as:
\begin{equation}
    n_\text{cl}(R) = \frac{\pi G\Sigma_\text{gas}\Sigma_\text{tot}}{12kT_\text{eff}}\approx 33 \left(\frac{\Pism}{P_{\odot}}\right)  T_{100}^{-1}\ \text{cm}^{-3},
    \label{eq:densidad_cl}
\end{equation}
 where $T_{100}$ is the temperature of the cloud gas in units of 100 K. The resulting values of $n_\text{cl}$ are a decreasing function of the galactocentric distance giving, for $T_{100}\sim1$,  a range of average densities between $\sim 200$ cm$^{-3}$ at 5 kpc and $\sim 10$ cm$^{-3}$ at 10 kpc, consistent with those observed for diffuse clouds \citep[see][]{Snow_McCall2006,Goldsmith2013,Neufeld+2024} and those modeled for CNM by \citet{Wolfire+2003}. 

Next, we discuss the conditions under which these CNM clouds, embedded within the warm neutral medium \citep[see][]{Kalberla_Haud2018,Park+2023},  can be transformed into molecular clouds and the rate at which this transformation occurs.

\section{Molecular clouds}
\label{sec:molecular_clouds}

The formation of molecular clouds (and eventually stars) can be viewed as a fragmentation process \citep{Franco_Cox86} in which the first step occurs when the central region of an initially cold atomic cloud is shielded from the external UV radiation field at wavelengths close to $\lambda\sim~10^{-5}$~cm. This shuts down its primary heat source, prevents the dissociation of H$_2$, and causes the eventual formation of a cooler and denser molecular cloud. As is already known, the external layers of molecular clouds are atomic transition regions, photodissociated regions (PDR), whose column densities protect the interior from the dissociating photons. Interstellar dust grains play three important roles in the fragmentation process. First, molecular hydrogen is formed on the grain surface lattice, second, small grains maintain the temperature of the gas through photoelectric heating, and third, the opacity of the small dust at the wavelengths of the dissociating Lyman and Werner bands near $\lambda \sim 10^{-5}$ cm reduces the photodissociation rate, protecting the newly formed H$_2$. Here we explore the role of Galactic parameters in controlling the formation process of the resulting molecular clouds.

\subsection{\textbf{Far ultraviolet dust extinction cross section:}}

Following the simplified scheme of \citet[][\citetalias{Franco_Cox86}]{Franco_Cox86}, the average mass density of dust grains in the ambient ISM, $\promedio{\rho_{\rm ISM,d}}$, is
\begin{equation}
 \promedio{\rho_{\rm ISM,d}} \approx n_\text{d} \rho_\text{d} V_\text{d},
\end{equation}
where $n_\text{d}$ is the number density of interstellar grains, $\rho_\text{d}$ is the internal mass density of a dust grain, and $V_\text{d}={4\pi a^3/3}$ is the volume of solid and homogeneous spherical grains with an average radius $a$ \citep[for porous or multilayer grains one can use an equivalent size parameter, see][]{Voshchinnikov+2006}. This average dust mass density in the ISM can be written in terms of metallicity $Z$, the dust-to-metal mass ratio $f_d$ (the fraction of heavy elements locked in the dust grains) and the mass density of the gas $\mu n_T$ (where $\mu = 2.27\times10^{-24}$ g is the mass per proton including He and $n_T=n_H+2n_{H_2}$ is the average number density of protons) as,
\begin{equation}  
 \promedio{\rho_{\rm ISM,d}} \approx Z~f_d~\mu~n_T.
\end{equation}
Then, the far-ultraviolet (FUV) extinction cross section per H atom can be written as
\begin{equation}
    \sigma_{\rm UV} \sim \pi a_{\rm UV}^2 \Qext\frac{n_d}{n_T}\approx \frac{3 \ \Qext\ Z\ f_d\ \mu}{4\ a_{\rm UV}\ \rho_d},
    \label{eq:sigma_uv}
\end{equation}
where $\Qext\sim 2.4$ is the extinction efficiency for compact, porous, or multilayer grains of effective size $a_{\rm UV}  \sim \lambda \sim 10^{-5}$ cm \citep[e.g.,][]{Voshchinnikov+2006}, and we take $\rho_d\sim 2.5$ gr cm$^{-3}$ as the average mass density within a grain (the internal densities of interplanetary dust grains range from 0.3 to 6.2 g cm$^{-3}$; \citealp{Love+1994}). With these values, the FUV cross section is
\begin{equation}
    \sigma_\text{UV} \approx 1.6\times 10^{-19} f_d \; Z \text{ cm}^2.
    \label{eq:sigma_uv_numerical}
\end{equation}
This linear dependence on $Z$ was first proposed by \citetalias{Franco_Cox86}, and was later independently adopted by several authors in the construction of PDR models \citep[e.g.,][]{Wolfire+2008,Krumholz+2008,Krumholz+2009,McKee+2010,Sternberg+2014,Maillard+2021}. However, a review of the observational restrictions imposed on the dependence on $Z$ and $f_d$, both of which are related to the dust-to-gas mass ratio, $D/G$, indicates that the dependence is not linear for metallicities $\leq (0.1-0.2)Z_{\odot}$, where it becomes much steeper \citep{Remy-Ruyer+2014,de_Vis+2019,Peroux_Howk2020}.

\subsubsection{The dust-to-gas mass ratio, $D/G$, in nearby galaxies:}\label{sec:D2GinNearbyGalaxies}

The relationship of $\sigma_\text{UV}$ with metallicity is linear, provided that the dust-to-gas mass ratio $D/G$ is also linear with $Z$. Several nearby galaxies show radial trends of $D/G$ that seem to be correlated with their metallicity gradients \citep[e.g.,][]{Issa+1990,Boissier+2004,Munoz-Mateos+2009,Mattsson_Andersen2012,Vilchez+2019}, and in a sample of 126 nearby galaxies, \citet{Remy-Ruyer+2014} derived a linear growth at metallicities higher than $12+
\log (O/H)\sim 8$, but the relationship changed to a cubic dependence for lower values. This break in the $Z$ dependence occurs at about $0.2 ~Z_{\odot}$, where $Z_{\odot}\approx 1.96\times 10^{-2}$ is the solar metallicity \citep{von_Steiger+2016}, and $12+\log (O/H)_{\odot}\sim 8.69$ is the value of the oxygen abundance in the solar circle \citep{Asplund+2009}. 

The nonlinear dependency at low metallicities was confirmed by \citet{de_Vis+2019}, who derived the values of $D/G$ as a function of $Z$ for some 500 galaxies from DustPedia \citep{Clark+2018} and other catalogs. They concluded that their data, which cover the same range of values of \citet{Remy-Ruyer+2014} but have a larger scatter, can be best fitted by a single power law with an exponent of 2.45.

More recently, \citet{Peroux_Howk2020} and \citet{Roman-Duval+2022b} made compilations of these and other studies, showing that the integrated values of the MW and Magellanic Clouds follow an almost linear dependence, with an integrated value for the MW of $(D/G)_{\rm MW} \sim 8\times 10^{-3}$. In addition, \citet{Clark+2023}\footnote{\citet{Clark+2023} found a correlation between $D/G$ and $\Sigma_{HI}$ for M31, M33 and the Magellanic Clouds, but given their flat (or non-existent) metallicity gradient, this is not related with $Z$ and may indicate dust growth in regions with high gas densities. The same argument applies to the dependence of elemental depletions on gas density fluctuations in the MW and Magellanic Clouds \citep[e.g.,][]{Roman-Duval+2022a}} and \citet{Hamanowicz+2024} obtained integrated values for other members of the Local Group (M31, M33. IC1613 and Sextans A), which also lie in the same almost linear track of the MW and Magellanic Clouds. The linear trend is also consistent with the results found by \citet{Draine+2007} and with those found by \citet{Wiseman+2017} for galaxies with a high redshift.  In addition, it should be noted that damped Lyman-$\alpha$ absorbers also display an intriguing linear relationship with metallicity.\footnote{\citet{De_Cia+2016} and \citet{De_Cia+2018} found an almost linear relationship of $D/G$ with $Z$ in damped Lyman-$\alpha$ absorbers covering two orders of magnitude in $Z$ that can be extended to those found in MW clouds. The trend and data are less scattered than those found in nearby galaxies, and a large fraction of their metallicity, probably most of it, seems to be locked in dust grains. }

Dust evolution models in disk galaxies can reproduce the main features and radial trends of $D/G$ versus $Z$ observed in nearby galaxies and indicate that dust formation can also occur in the ISM \citep[e.g.,][]{Mattsson_Andersen2012,Feldmann+2015,Zhukovska+2016,Hou+2019,Aoyama+2020}. Together, these theoretical and observational results indicate that the linear dependence can be used for metallicities above $\sim 0.2~Z_\odot$, but the relationship is much steeper, approximately cubic, below this value.

\subsubsection{The dust-to-metal mass ratio, $f_d$, in nearby galaxies:}

The equilibrium values of the dust-to-metal mass ratio in the ISM, defined as $f_d=D/M=(D/G)/Z$, depend on the formation, growth, and destruction mechanisms of the grains that operate in the ISM. Thus, for metallicities above $0.2~Z_\odot$, where $D/G$ is linear with $Z$, it is expected that $f_d$ should reach a constant value after the dust lifecycle reaches equilibrium within a given galactic system. 

Data compilation of 15 nearby galaxies by \citet{Mattsson_Andersen2012} display radial gradients in both $D/G$ and $f_d$, but the trends in $f_d$ are much flatter and in some cases are almost constant. These shallow trends are approximately reproduced by numerical models of chemical enrichment and dust evolution in gaseous disks of late-type galaxies \citep[e.g.,][]{Mattsson_Andersen2012, Aoyama+2020}, and these and other models indicate that as the amount of dust-forming elements decreases by depletion in the gas phase, dust growth slows and $f_d$ tends to a constant value (see \citealp{Chiang+2021} and references there in). Interestingly, the constancy of $f_d$ has been found even in metal-poor systems with $Z\sim 10^{-2}Z_\odot$ by \citet{Zafar_Watson2013}.

\citet{Chiang+2021} performed a detailed study of the dust-to-metal ratio in five nearby galaxies and found that they are about constant within each galaxy, with equilibrium values ranging from 0.4 to 0.58. Their data have less scatter compared to other similar studies (see Figures 5 and 6 of \citealp{Chiang+2021}), and are also consistent with a linear dependence of $D/G$ on $Z$ at metallicities greater than 0.2 $Z_\odot$. For the metallicity in the solar neighborhood, their results and the linear fit of \citet{Remy-Ruyer+2014} give an average equilibrium value close to $f_d\sim 0.6$. Thus, using Eqs.~(\ref{eq:sigma_uv}) and (\ref{eq:sigma_uv_numerical}) with $f_\text{d}\sim 0.6$, we recover the cross-sectional value of \citetalias{Franco_Cox86}, 
\begin{equation}
\sigma_{\rm UV} \approx 1.9 \times 10^{-21}(Z/Z_{\odot})\ \text{cm}^2.
\end{equation}
But its validity is now limited to metallicities greater than $\sim 0.2~Z_{\odot}$ and the value in the solar circle is similar to those derived from observations in the diffuse medium with an optical extinction of $A_v/E(B-V)\approx 3.1$, and to those used in PDR models \citep[e.g.,][]{Bohlin+1978,Draine_Bertoldi1996,Rollig+2007,Krumholz+2008,Rollig+2013,Maillard+2021,Shull+2021}. 

\subsubsection{The H$_2$ formation rate as a function of $\Pism$:}

One of the first molecules to be formed in a nascent molecular cloud is H$_2$, less abundant molecules are formed later. Its formation rate on the surface of dust grains is $n_{\rm HI} n_T R_f$, so the formation rate per hydrogen atom and unit time is 
\begin{equation}
F=n_{T}R_f, 
\label{eq:FormationRatePerH}
\end{equation}
where $n_{\rm HI}$ is the abundance of atomic hydrogen, $n_T=n_{HI}+2n_{H_2}$ is, as above, the average total abundance of protons, and the rate coefficient for the formation of H$_2$ on grains is \citep{Hollenbach+1979}
\begin{equation}
R_f=\frac{n_d \sigma_d}{n_T} \promedio{Sv} f_a ,
\end{equation}
where $n_d$ is the volumetric density of dust grains in the cloud, $\sigma_d$ is the grain cross section for collisions with atomic hydrogen, $f_a$ is the fraction of H atoms that stick to the grain to form H$_2$, and $\promedio{Sv}\approx 7\times 10^4 \ T_{100}^{1/2}$ cm s$^{-1}$ is the average velocity of the incoming H atoms.  For CNM clouds, which have low $T$ and the expected value of $f_a$ tends to 1, the approximation of \citet{Hollenbach+1979} and the results of \citet{Jura75} from UV observations give a rate coefficient of $R_f\sim 3\times 10^{-17}\ T_{100}^{1/2}\ \text{cm}^3\ \text{s}^{-1}$. 

For pressure bounded clouds, $n_T$ is proportional to $\Pism/T_{100}$ (eq.~\ref{eq:densidad_cl}), and the formation rate per hydrogen atom (Eq.~\ref{eq:FormationRatePerH}) can be approximated by
\begin{equation}
    \begin{aligned}        
    F\sim & 10^{-15}\left( \frac{ \Pism}{P_\odot}\right)\ T_{100}^{-1/2}\ s^{-1} \\
    & \approx 3.16 \times 10^{-8}  \left( \frac{ \Pism}{P_\odot}\right)\ T_{100}^{-1/2} \ \text{yr}^{-1}.
    \end{aligned}
\label{eq:H2formation_rate}
\end{equation}
Thus, the transformation of HI into H$_2$ in the inner parts of a diffuse cloud is directly controlled by ambient pressure. In addition, given that dust opacity depends on $Z$, the net result is that the formation of molecular clouds depends on both $\Pism$ and $Z$ and the process is faster and more efficient in the inner regions of galaxies, where both pressure and metallicity are higher.

\subsection{\textbf{Atomic to molecular transition regions}}

Details of the PDR region that separates the diffuse medium from the molecular gas have been explored for the last five decades with a number of different models and methods \citep[e.g.,][]{Hollenbach+1971,Glassgold_Lange_1974, Jura1975a,Jura1975b,Federman+1979,de_Jong+1980,Tielens_Hollenbach1985,Elmegreen1989,Abgrall+1992,Draine_Bertoldi1996,Browning+2003,Rollig+2007,Krumholz+2008,McKee+2010,Sternberg+2014,Gong+17,Maillard+2021}. The impinging FUV radiation field at $\lambda \geq 916 \text{ Å}$ is usually expressed in units of the local interstellar field described by \citet{Habing1968}, $4\times 10^{-17}$ erg cm$^{-3}\text{ Å}^{-1}$, or in terms of the updated value derived by \citet{Draine+1978}, which is 1.71 higher. As these photons enter a certain distance from the external edge of the cloud, their intensity decreases due to the opacity of the dust, and the corresponding photodissociation rate in unidirectional plane-parallel approximations drops as $\exp\ (-\tau_d)$, where $\tau_d=\int \sigma_{\rm UV}n_Tdx$ is the optical depth. As molecules begin to form, they absorb FUV photons and increase opacity. 

For isotropically illuminated models, the attenuation of the radiation field at a given point from the external cloud boundary is smaller than in unidirectional models because the incoming photon flux comes from different directions, and this is further complicated when dust scattering is included \citep[e.g.,][]{Flannery+1980,Rollig+2007}. Given that molecular clouds, either individual or in cloud complexes, have irregular or filamentary shapes without preferred geometries, unidirectional and plane-parallel configurations have been used as a reasonable approach to the problem \citep[see, e. g.,][]{Gong+17}. 

With this approach, \citetalias{Franco_Cox86} proposed that a proxy for the column density to protect the center of the cloud from dissociating radiation in the solar neighborhood can be approximated by the condition of $\tau_d \sim 1$,
\begin{equation}
  N_d\approx \sigma_{\rm UV}^{-1}\approx 5 \times 10^{20} \left( \frac{Z_{\odot}}{Z}\right) \text{cm}^{-2}. 
\end{equation}
Or, equivalently, diffuse clouds with a radius larger than
\begin{equation}
    r_{\tau}= \frac{N_{d}}{n_\text{cl}}
    \approx 3 \left(\frac{Z_\odot}{Z}\right)\left(\frac{P_\odot}{\Pism}\right) T_{100}\ \text{pc}, 
\end{equation}
become molecular at their centers. These threshold values are more than satisfied by the amount of mass gathered in convergent gas flows, instabilities, or shocks in spiral arms \citep[e.g.,][]{Jog_Solomon1984,Elmegreen1987,Hartmann+01,Kim_Ostriker02,Franco+2002,Schaye2004,Zamora-Aviles+19}. As the amount of molecules accumulates in its interior, the original cold diffuse cloud readjusts its original structure to a centrally condensed configuration: becoming denser and cooler at the core, while remaining atomic and warmer at the peripheral layers, but in pressure equilibrium with the ambient medium. Also, given that the central cloud temperatures are lower, the corresponding velocity dispersions should increase toward the center of the cloud \citep[see][]{Ballesteros-Paredes+11a}. 

This simple proxy provides remarkable results. Observations in the ultraviolet indicate that the threshold for molecular cloud formation in the MW is achieved when the ratio of molecular to total column density is $2N_{H_2}/N_H\approx 0.1$, where $N_H=N(HI)+2N(H_2)$, resulting in a transition region with a total column density of gas of approximately $\sim 5\times 10^{20}$ cm$^{-2}$ \citep[e.g.,][]{Savage+1977,Bohlin+1978,Federman+1979,Shull+2021,Park+2023}, similar to Eq. (23) and equivalent to a mass layer of $\sim 5\ M_\odot \ \text{pc}^{-2}$. The proxy is also consistent with the threshold of the HI-H$_2$ transition region derived from FUV observations in Magellanic Clouds \citep{Tumlinson+2002,Browning+2003,Shull+2021}, where the average integrated metallicity values are $Z_{\rm LMC}~\sim~0.5~Z_\odot $ and $Z_{\rm SMC}~\sim~0.2~Z_\odot $ \citep[e.g.,][]{Roman-Duval+2022a} and is also consistent with the FUV observations of H$_2$ in starburst galaxies \citep{Hoopes+2004}.

On the other hand, as noticed by \citet[][see also \citealp{Wong_Blitz2002}]{Martin_Kennicutt01} in a sample of 32 galaxies, disks become dominated by HI gas when the gas density drops below $5-10\ M_\odot\ \text{pc}^{-2}$. Later, \citet{Bigiel+2008} showed that for 18 nearby spirals the transition from atomic to molecular gas can occur around $\sim10 \  M_\odot\ \text{pc}^{-2}$ corresponding to a column density of $\sim10^{21}$ cm$^{-2}$ (see their Figure 8). This value coincides with the proxy along the entire cloud, namely $2N_d$,
\begin{equation}
2N_d=2\ \sigma_{\rm UV}^{-1}\approx 10^{21} \left( \frac{Z_{\odot}}{Z}\right) \text{cm}^{-2}, 
\ \ \ 
 2\Sigma_d\approx 10 \left( \frac{Z_{\odot}}{Z}\right) {M}_\odot\ \text{pc}^{-2},
 \label{eq:opacity_mass_col_density}
\end{equation}
and to that required for the appearance of local dynamical instabilities that lead to a transition to cold ISM phases, as derived by \citet{Schaye2004}. Therefore, despite its simplicity, the $\tau_d\sim 1$ condition provides a reasonable approximation to the observed transition values in the Local Group and nearby spirals. 

One may ask how such a simple criterion compares with the results of elaborate PDR models. \citet{Shull+2021} made an analytical match of their observational FUV results for the transition region using detailed stationary PDR models from \citet{Draine_Bertoldi1996} (these include dust opacity, self-shielding of thousands of H$_2$ lines, and the effects of line overlap). They found that their observations are explained by an optical depth of dust of $\tau_d = 1.013$. To complement this result, we derive the corresponding equilibrium volume density of H$_2$ at the location of the transition point using the formation rate described above and the approximation to the photodissociation rate given by \citet{Draine_Bertoldi1996}, 
\begin{equation}
  G_{\rm diss} \approx f_{\rm shield}\ \text{exp}(-\tau_d)\ G_d,
\end{equation}
where $f_{\rm shield}= 3.16 \times 10^{10}N(H_2)^{-3/4}$ is the shielding function and $G_d$ is the photodissociation rate at the external edge of the cloud. For the solar neighborhood, the ambient FUV radiation field derived by \citet{Habing1968} gives a photodissociation rate at the external edge of the cloud, $G_\odot\sim 6\times 10^{-11}$ s$^{-1}$ \citep{Federman+1979} and, as usual, we write $G_d=\chi G_\odot$ to allow for different intensities of the FUV field (the updated value of \citet{Draine+1978} corresponds to $\chi= 1.71$).

With these numbers, the equilibrium molecular-to-atomic ratio at $\tau_d\sim 1$ is
\begin{equation}
\begin{aligned}
\frac{n_{H_2}}{n_{HI}}= &\frac{n_TR_f}{f_{\rm shield}\ \exp(-\tau_d)G_d} \\
& \approx 0.3 \left( \frac{ \chi}{1.71}\right)^{-1} \left(\frac{ \Pism}{P_\odot}\right)T_{100}^{-1/2},
\end{aligned}
\end{equation}
and the resulting fraction of H$_2$ at the point of transition is
\begin{equation}
  \frac{n_{H_2}}{n_T}\sim~0.2~\left( \frac{ \chi}{1.71}\right)^{-1}\left( \frac{ \Pism}{P_\odot}\right)T_{100}^{-1/2}.
\end{equation}
Therefore, as expected, most of the gas in the HI-H$_2$ transition layer is atomic and, aside from its dependence on the FUV radiation field, the fractional abundance of H$_2$ in the cloud interior is regulated by the pressure of the system. 

Several observational results, ranging from radio waves to the FUV in the MW and in nearby spirals and irregulars, indicate that there is a coincidence in the threshold to the existence of molecular clouds and the formation of stars, occurring at gas column densities in the range $\sim 5-15\ M_\odot \ \text{pc}^{-2}$ \citep[e.g.,][]{Skillman1987,Kennicutt1979,Martin_Kennicutt01,Wong_Blitz2002,Bigiel+2008}. However, we note that, despite similarities in numerical values, they refer to fundamentally different issues. The transition region in the FUV refers to the column density of the mainly atomic hydrogen layers that surround any molecular cloud, $\Sigma_{d}$, while the threshold for star formation refers to the surface density of the molecular hydrogen mass, $\Sigma_{H_2}$ inside the clouds. 

\subsection{\textbf{Radial threshold conditions}}

Using the radial gradient of $O/H$ obtained by \citet{Arellano-Cordova+2021} as the metallicity tracer for the MW, $12+\log (O/H)\approx 8.84 - 0.042 \ R$, the radial variation between 5 and 13 kpc can be approximated by
\begin{equation}
Z \approx Z_{\odot}~\text{exp} \left(\frac{R_{\odot}-R}{10\ \text{kpc}}\right) .
\end{equation} 
Note that our selected $(O/H)$ value for the solar circle in \S\ref{sec:D2GinNearbyGalaxies} is smaller than the one used by \citet{Arellano-Cordova+2021} but is in better agreement with their actual fit at 8 kpc. The resulting threshold mass column density variation along the disk of the MW is then (see Eq. \ref{eq:opacity_mass_col_density}),
\begin{equation}
\Sigma_{d} \approx 5\ \text{exp} \left(\frac{R-R_{\odot}}{10\ \text{kpc}}\right)M_\odot \ \text{pc}^{-2}.
\end{equation}
For locations close to the molecular ring at 4 kpc, the corresponding pressure is about six times higher, and the metallicity is about 1.5 times higher than those at the solar circle, and cold clouds with mass surface densities as small as 3 $M_\odot\ \text{pc}^{-2}$ and radii as small as $\sim$ 0.3 pc fulfill the requirement to become molecular. 

Given that the population of molecular clouds is controlled by metallicity and system pressure, the obvious and well-known result is that most of the molecular mass should be concentrated in the inner parts of the gaseous disk. Another consequence of the larger pressure values within the solar circle, which is also obvious, is that the average densities of individual molecular clouds should be higher, in accordance with those derived for individual clouds in the compilation of \citet{Miville-Deschenes+17}. 

\subsection{\textbf{Pressure-regulated formation of molecular gas along the disk of the MW}}

Recent emission and absorption studies of atomic hydrogen show that, in addition to the already known cold and warm stable phases, the third and unstable "lukewarm" phase is also prevalent in the disk of the MW \citep{Kalberla_Haud2018,Park+2023,Gupta+2025}. The relative fraction of these three phases is somewhat similar and, assuming that the actual precursor of molecular clouds is the CNM phase, only about a third of $\Sigma_{HI}\approx \ 5 \ M_\odot \ \text{pc}^{-2}$ (see Section \ref{sec:MW_radial_distr}) is available to form molecular clouds. The resulting rate of molecular gas formation per unit surface in the MW disk, $\dot{\Sigma}^{+}_{\rm mol}$, is simply given by the product of $\Sigma_{HI}/3\approx 1.7\ M_\odot \ \text{pc}^{-2}$ with the formation rate of H$_2$ per hydrogen atom, Eq. (\ref{eq:H2formation_rate}),
\begin{equation}
\begin{aligned}
\dot{\Sigma}^{+}_{\rm mol}=\frac{\Sigma_{HI} F}{3} 
&\sim 1.7\times 10^{-9}  \left( \frac{ \Pism}{P_\odot}\right)\ T_{100}^{-1/2} \ M_\odot\ \text{kpc}^{-2}\text{s}^{-1} \\
& \approx 5\times 10^{-2}\ \text{exp} \left(\frac{R_{\odot}-R}{2\ \text{kpc}}\right)T_{100}^{-1/2}\ M_\odot\ \text{kpc}^{-2}\text{yr}^{-1}.
\end{aligned}
\end{equation}
This H$_2$ mass formation rate can be termed pressure-regulated and assuming that the cycling of the gas components of the disk is in dynamical equilibrium, it should be equal to the rate of destruction of the molecular gas per unit surface, $\dot{\Sigma}^{-}_{\rm mol}$. 

The destruction rate is the sum of the transformation of the molecular mass into stars and the evaporation of the parental clouds due to the injection of energy from newly formed stars. Thus, the cycling between atomic and molecular gases owing to stellar feedback in equilibrium is also regulated by pressure, which supports the self-regulated model of \citet{Ostriker+10}. In the solar circle, then, the rate of molecular cloud destruction should be about $\sim 5\times 10^{-2} \ M_\odot\ \text{kpc}^{-2}\text{yr}^{-1}$, and its implications are discussed in the next section. Note that in the hypothetical case where no stars are formed, $\dot{\Sigma}^{-}_{\rm mol}=0$, the molecularization of the disk occurs inside-out and on a time scale of about $\Sigma_{HI}/\dot{\Sigma}^{+}_{\rm mol}\sim 10^8\left(P_\odot/{\Pism}\right)T_{100}^{1/2}$ yr.

\section{Self Gravitating Clouds and Star Formation}
\label{sec:selg_grav_clouds_and_sfr}

The onset of self-gravity is achieved when the gravitational energy of an initially pressure-confined cloud is similar to its internal energy $GM_\text{cl}^2/R_\text{cl}\sim P_{\rm ISM}V_\text{cl}$, and the corresponding threshold mass surface density across the cloud is
 \begin{equation}  
 \Sigma_{\rm sg} \approx \left(\frac{4\Pism}{\eta G}\right)^{1/2},
\end{equation}
where $V_\text{cl}=\eta R_\text{cl}^3$ is the volume of the cloud, and the geometric factor for uniform spherical clouds is $\eta_s=4\pi /3$. Note that this surface mass density criterion is equivalent to the one derived from the Jeans length criterion\footnote{Except for a factor close to unity, this result is equal to the product of twice the Jeans length for spherical clouds $L_J=[15kT/(4\pi \mu G\rho_{cl})]^{1/2}$ and the average mass density of our pressure-bounded clouds  $\rho_\text{cl}~=~\mu~P_\text{ISM}/(6~kT)$, $2L_J\rho_{cl}=\Sigma_J=[20\Pism/(8\pi G)]^{1/2}$.}. For the solar circle, the threshold value is
\begin{equation}  
 \Sigma_{\rm sg, \odot}\sim 30 \left(\frac{\eta_s}{\eta}\right)^{1/2} M_{\odot} \ {\rm pc}^{-2},
\end{equation}
and its radial distribution along the disk is
\begin{equation}  
 \Sigma_{\rm sg} =  \Sigma_{\rm sg, \odot} \left(\frac{P_\text{ISM}}{P_{\odot}}\right)^{1/2}\sim 30\ \left(\frac{\eta_s}{\eta}\right)^{1/2} \text{exp} \left(\frac{R_{\odot}-R}{4\ \text{kpc}}\right) M_{\odot} \ {\rm pc}^{-2}.
\end{equation}
Thus, the formation of self-gravitating clouds also depends on the pressure of the system, and the required surface mass density for the transformation of a molecular cloud into a self-gravitating entity can be collected via gravitational or magnetic instabilities or by the convergence of large-scale streams of gas \citep[][]{Ballesteros-Paredes+99a, Ballesteros-Paredes+99b, Hartmann+01, Franco+2002,Lee+2004,Vazquez-Semadeni+07, Heitsch_Hartmann08}. At some moment in the assembly process, the gravitational energy becomes greater than the sum of the internal energies of the cloud, and its interior decouples from the pressure of the system. 

\begin{figure}
    \centering
    \includegraphics[width=\columnwidth]{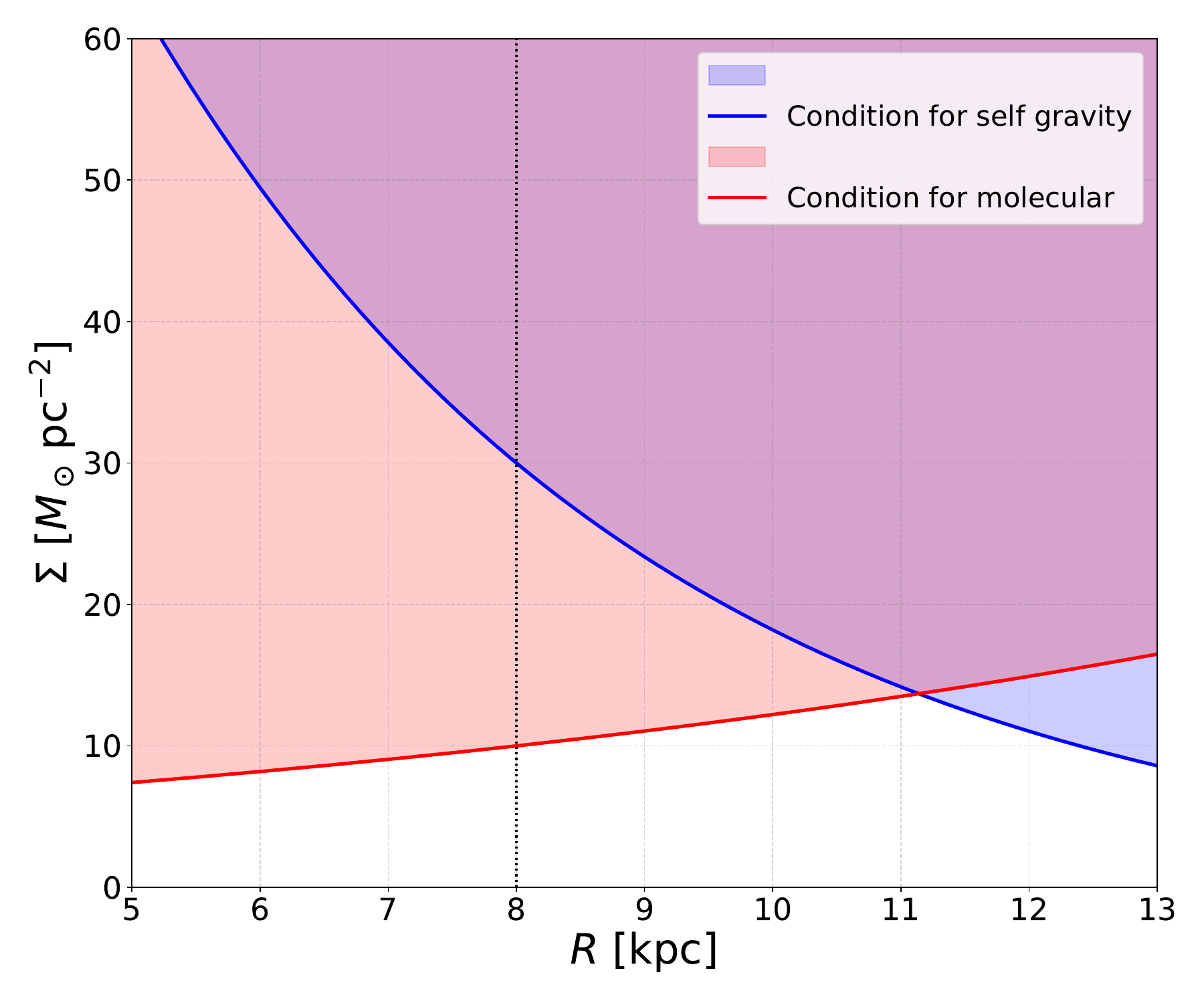}
    \caption{The threshold mass surface density across the whole cloud to become molecular, $2\Sigma_d$ (red line), and the threshold condition to become self-gravitating, $\Sigma_{sg}$ (blue line). Clouds with gas mass surface densities above the red line are molecular (shaded in pink), while those above the blue line are self-gravitating (shaded in blue).}
    \label{fig:Thresholds_molecular_and_self-gravity}
\end{figure}

Figure \ref{fig:Thresholds_molecular_and_self-gravity} shows the molecular and self-gravity threshold criteria, and they become equal at a galactocentric distance of about $\sim$11 kpc. This distance marks the location inside which the star-forming clouds follow a two-step evolution. First, they become molecular, and then, after contracting or gathering mass from the external medium, their column density increases, and they become self-gravitating, starting a track towards the formation of stars, typically in a hierarchically and nonmonolitic mode \citep[e.g., ][]{Ballesteros-Paredes+11a, Vazquez-Semadeni+19}. The figure also shows, as noted by \citet{Evans+2021}, that even when molecular clouds in the inner galaxy have larger surface mass densities, they are not necessarily self-gravitating. 

For locations outside 11 kpc, the decrease in both pressure and metallicity results in less dense and less opaque clouds. As a consequence, the population of molecular clouds should drop, but the existing ones first became self-gravitating, when they reached total gas column densities of about 12 M$_{\odot}$ pc$^{-2}$, and then became molecular after reaching the opacity criterion. In these external regions, then, the evolution of clouds is different, and the star formation process can also occur in small molecular clouds and even in mainly atomic clouds. This is in line with the results of the high-resolution survey of molecular clumps in the outer MW by \citet{Urquhart+2025}, where the fraction of low-mass star-forming clumps is higher than in the inner MW. Also, this possibility may be at work in the formation of intergalactic $HII$ regions \citep[see][]{Ryan-Weber+2004}.

In a study with more than 8100 molecular clouds in the Milky Way, \citet{Miville-Deschenes+2017} obtained the physical properties and radial trends of individual clouds located as far as 20 kpc from the Galactic center. They found that the densities (both volumetric and per unit area), as well as the velocity dispersions, are substantially larger in the inner galaxy than in the outer parts, as expected from the results of the previous section. 

The median of mass surface values for individual clouds located between 6 and 10 kpc follows the trend of interstellar pressure, and coincidentally scales with one-half of the value of the threshold for self-gravity at the solar circle, $\Sigma_{\rm sg,\odot }/2$ (see their Figure 9), 
\begin{equation}  
 \Sigma_{\rm median} \sim  0.5\ \Sigma_{\rm sg,\odot} \left(\frac{\Pism}{  P_{\odot}}\right) \approx15 \ \text{exp} \left(\frac{R_{\odot}-R}{2\ \text{kpc}}\right) M_{\odot} \ {\rm pc}^{-2}.
\end{equation}
This may be another indication that the physical properties of the molecular cloud population are controlled by the pressure of the system, and, given that the median value implies that half of the clouds are below such a value, a good fraction of the clouds in the compilation of \citet{Miville-Deschenes+17} are not self-gravitating.

In order to compare this result with our expected total surface mass density of H$_2$ at the threshold of self-gravity, we have to subtract the contribution of the atomic layers that surround the central molecular gas. Then, the total H$_2$ surface mass density across a cloud at the threshold of self-gravity is $\Sigma_{t,\rm mol}=\Sigma_{\rm sg}-2\Sigma_d$,

\begin{equation}
\begin{aligned}
    \Sigma_{t,\rm mol} & \approx \\&\left[30\ \left(\frac{\eta_s}{\eta}\right)^{1/2} \text{exp} \left(\frac{R_{\odot}-R}{4\ \text{kpc}}\right) - 10\ \text{exp} \left(\frac{R-R_{\odot}}{10\ \text{kpc}}\right) \right]  \ 
    M_{\odot} \ {\rm pc}^{-2}.
\end{aligned}
\end{equation}
Figure 3 compares Eq. (36) with the median values of the compilation of \citet{Miville-Deschenes+17}. It shows that molecular clouds inside 6 kpc and outside 10 kpc tend to be self-gravitating but, as expected, a good fraction of molecular clouds between 6 and 10 kpc are not gravitationally bound. This is also in line with the fact that a large number of the cloud population display virial parameters well above the value expected for gravitationally bound systems, as pointed out by \citet{Evans+2021}. 

Also note that our criterion in the solar neighborhood corresponds to about $\Sigma_{t,\rm mol}(R_\odot)\sim 20 \ (\eta_s / \eta)^{1/2}$  M$_{\odot}$ pc$^{-2}$. For $(\eta_s / \eta)^{1/2}=1$, it is higher than the range of observed threshold molecular column densities for star formation, $\sim 5-15\ M_\odot \ \text{pc}^{-2}$ \citep[e.g.,][]{Skillman1987,Martin_Kennicutt01,Wong_Blitz2002,Bigiel+2008}, but the observational values can be matched if $(\eta_s / \eta)^{1/2}\sim (0.3-0.7)$, indicating that clouds are possibly filamentary. 

\begin{figure}
    \centering
    \includegraphics[width=\columnwidth]{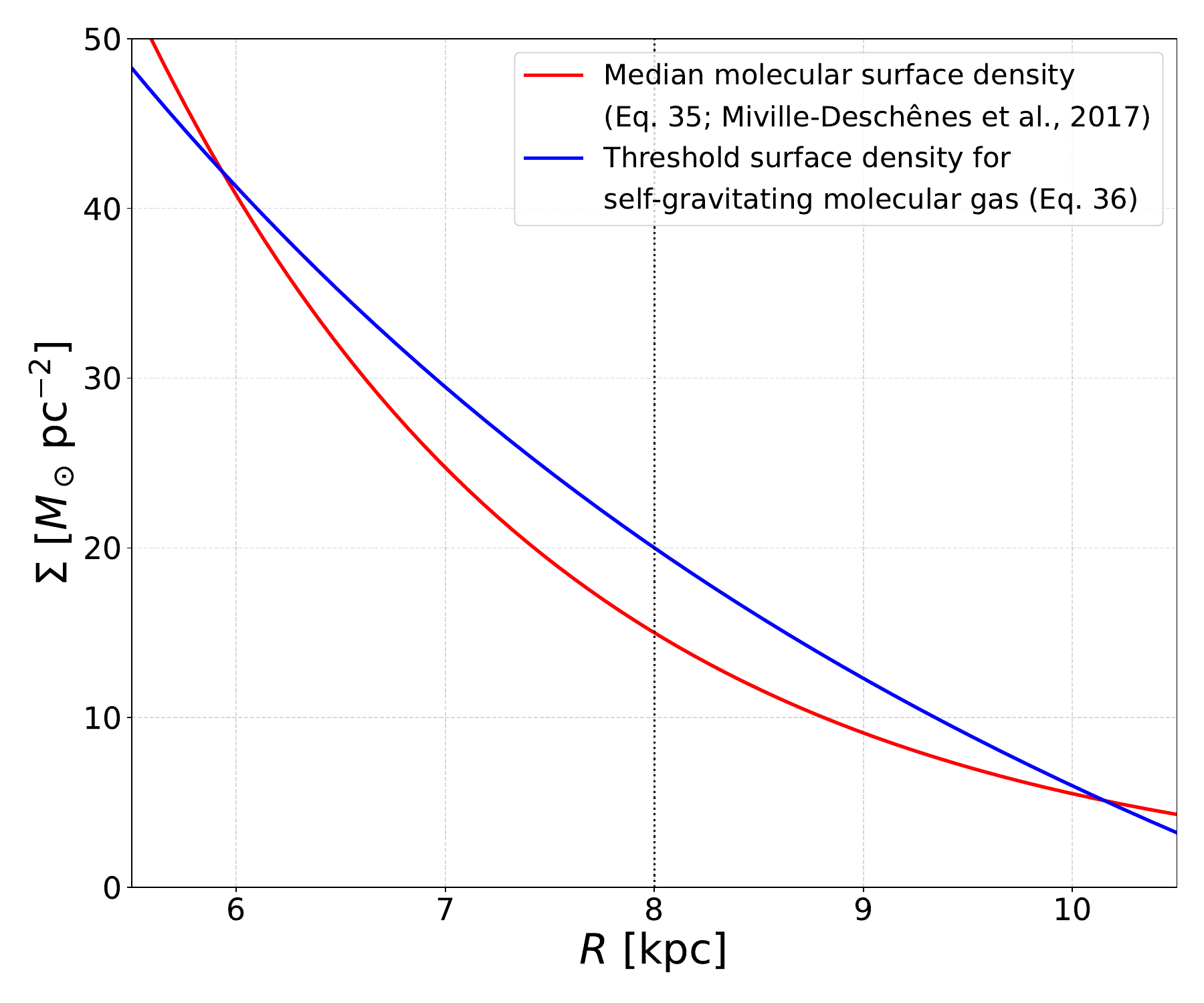}
    
    \caption{The median mass surface density  between 6 and 10 kpc from the compilation of \citet{Miville-Deschenes+2017}, as approximated in Ec. (35) (red line). The total H$_2$ surface mass density threshold condition to become self-gravitating of our model, from Ec. (36) (blue line).}
    \label{Median-MolSG}
\end{figure}

\subsection{\textbf{Arm-interarm pressure differences}}

Concerning the role played by spiral arms\footnote{See \citet{Khoperskov+2013} and \citet{Khoperskov+16} for the formation of molecules and molecular clouds along the arms}, there are numerous studies exploring their role in star formation, including large-scale magnetic instabilities, gravitational fragmentation within the arms, and the formation of dust lanes due to the arm-interarm interaction \citep[e.g.,][and references therein]{Franco+2002, kim+02,Bonnell+13,Renaud+2013,Khoperskov+2013,Elmegreen_Elmegreen2019}. For our purposes, it is sufficient to say that the pressure inside the arm increases because the total mass and gas surface densities are higher than in the inter-arm regions. For example, in the case of arms located at 4.5 and 6.5 kpc in the MW, \citet{Marasco+2017} found that the surface mass densities of both HI and H$_2$ are factors approximately 1.5 to 2 above the average values at these locations, while external galaxies show similar H$_2$ contrasts \citep{Querejeta+2024}. The models of \citet{Gomez_Cox2004}, \citet{Khoperskov+2013} and \citet{Mata-Chavez+2019} for spiral arms in normal spirals indicate that the total mass surface density increases by a factor of about two or three. Thus, the product $\Sigma_g\Sigma_{tot}$ within the arms is about a factor of 4 above the interam regions, and the arm pressure increases by the same factor. This also increases the cloud densities by the same factor, directly increasing the rate at which molecules are formed within the arms and leading to the prevalence of dense molecular clouds along the arms. 

Cloud agglomeration and instabilities also result in most giant molecular clouds located inside the arms. In addition, the radiative shock front associated with the passage of the arm compresses even further the already dense molecular clouds and can trigger the rapid formation of stars, making these locations preferential sites for star formation. Our scheme indicates that the surface density contrast of the arm-interarm H$_2$ gas should be proportional to the pressure contrast. The H$_2$ surface density data of \citet{Marasco+2017}, displayed in their Figure 10, show that the arm-interarm surface density contrast is indeed between 3 and 4, close to the one expected from our pressure-regulated scheme.

An additional issue is that the arm shock front in a magnetized medium produces a hydraulic jump, sending gas to high latitudes \citep{Martos_Cox1998} and can even induce star formation outside the main gaseous disk \citep{Martos+1999}. The HI data of \citet{Marasco+2017} show an increase in the gas scale height (possibly due to the passage of the arms), and this increase is offset by about 1 kpc from the locations of the gas density peaks (see their figures 11, 14 and 15). A recent study with 3D dust maps by \citet{Kormann+25} has discovered large wave-like gas structures in the solar neighborhood. These are kiloparsec-sized atomic superclouds with corrugations locating the gas above and below the midplane. It is unclear whether they are associated or not with the passage of any spiral arm, but they resemble the outcome of the Parker instability in 3D \citep[see][]{Franco+2002}.

\subsection{\textbf{Pressure-regulated star formation per unit area in the disk of the MW}}
\label{sec:radia_SFR_distr}

Observational studies show that compared to other spirals, where interstellar pressure values can reach log$(\Pism /k) \sim 8.5$ \citep[e.g.,][]{Blitz_Rosolowsky_2006,Fisher+2019,Barrera-Ballesteros+21,Sun+23,Ellison+24}, the MW disk can be considered a low-pressure system with modest values up to log$(\Pism /k) \sim 5$. These works also show that the surface density of star formation increases approximately linearly with pressure for log$(\Pism /k) \leq 5.5$ but the trend becomes shallower at higher pressures \citep[e.g.,][]{Fisher+2019,Barrera-Ballesteros+21,Ellison+24}. This linear relation with $\Pism$ has been deduced by models of feedback-regulated star formation, as discussed by \citet{Ostriker+10} and \citet{Ostriker_Kim22}. Also, as stated in Eq. (13), we can infer it from the relationship between pressure and molecular gas surface density. Here we show that the linear relation between pressure and surface density of star formation naturally arises from basic principles. 

The star formation rate per unit area should be proportional to $\Sigma_{\rm sg}/ t_{\rm ff}$, where $t_{\rm ff}$~=~$(32G\rho_\text{cl} / 3\pi)^{-1/2}$ is the free fall time and $\rho_\text{cl}$ is the average volume mass density of a cloud bounded by pressure and at the beginning of self-gravity. This average volume mass density is as before $\rho_\text{cl}~=~\mu~P_\text{ISM}/(6~kT_\text{eff})$, and the free fall time for the parameters of our system is
\begin{equation}
    t_\text{ff}\sim 7 \left(\frac{P_\text{ISM}}{P_\odot \ }\right)^{1/2} T_{100}^{-1/2}\text{ Myr}.
\end{equation}
Theoretical and observational work have suggested that the entire life cycle of star-forming molecular clouds, from the time of their formation out from the CNM to their disruption by stellar feedback, lasts some $\sim10-30$~ Myr and between $10-15$~ Myr for recent extragalactic estimates \citep[see, e.g., ][]{Blitz+07, Fukui_Kawamura10, Colin+13, Vazquez-Semadeni+19, Chevance+20b, Chevance+23}. The corresponding timescale for individual cloud clumps, or small clouds in the solar neighborhood, is smaller, on the order of $5-10$~Myr \citep[][]{Leisawitz+89, Ballesteros-Paredes+99b, Hartmann+01, Ballesteros-Paredes_Hartmann07}, and this range of values is expected for self-gravitating cores. Assuming $t_{\rm ff}$ as the lifetime of our self-gravitating structures, the resulting star formation rate per unit surface can be written as
\begin{equation} 
\Sigma_{\rm SFR} \approx  \delta_a~\frac{\Sigma_{\rm sg}}{t_{\rm ff}} \approx  \delta_a \ 4\ \left(\frac {\Pism}{P_\odot }\right) T_{100}^{-1/2}\ \text {M}_\odot~\text {kpc}^{-2} \text {yr}^{-1},
\end{equation} 
where the proportionality factor $\delta_a$ corresponds to the fractional area subtended by star formation sites per kpc$^{-2}$. 

As stated in Sect. (2.2), from the value of Eq. (13) at the solar circle one readily gets $\delta_a \sim 4\times 10^{-4}\ T_{100}^{1/2}$ which, as we see below, represents a lower bound to $\Sigma_{\rm SFR}(R_\odot)$. The upper bound can be estimated using the available molecular cloud data. The surface mass density of molecular gas in the solar circle is, as stated before, $\Sigma_{H_2} (R_{\odot})\approx 1.5\ M_\odot \ \text{pc}^{-2}$, while the average mass of individual molecular clouds in Table 2 of \citet{Miville-Deschenes+17} is about $1.5\times 10^5 M_\odot$. Using the mass-radius relationship of the survey performed by \citet{Roman-Duval+10}, $M_\text{cl}\approx 228\  R_\text{cl}^{2.36}$ $M_\odot$, where $R_\text{cl}$ is the radius of the molecular cloud in pc, a cloud with $1.5\times 10^5 M_\odot$ should have a radius of about 15~pc\footnote{This a factor of 2 less than the average cloud size of \citet{Miville-Deschenes+17}. The high-resolution survey performed by \citet{Roman-Duval+10} provides better estimates of the clouds' sizes. Their Table~1 shows that clouds with masses around $10^5~M_\odot$ can have radii as small as 4 pc, but most of them are close to 15 pc. Thus, the average values of \citet{Miville-Deschenes+17} tend to overestimate cloud sizes and underestimate cloud densities.}. Thus, the effective area covered by each cloud complex, if round, is $\sim 7 \times 10^{-4}$ kpc$^2$. 

The detailed census of the main molecular structures located within 2 kpc of the Sun performed by \citet{Spilker+21} shows that there are about six large molecular clouds per square kpc\footnote{The 3D dust maps recently studied by \citet{Kormann+25} display the presence of large wavelike gas structures in the solar neighborhood. These are corrugated, kiloparsec-sized atomic superclouds, which appear to be precursors for future giant molecular cloud complexes.}. Then the fraction of the area covered in kpc$^2$ by these six clouds is $\sim 4 \times 10^{-3}$. Now, as already shown in Figure 3, a large fraction of molecular clouds between 6 and 10 kpc are not self-gravitating, and \citet{Evans+21,Evans+22} found that only 17\% to 19\% of the CO mass in the MW is gravitationally bounded. Then, as a first approximation, we assume that the cover area of the self-gravitating structures is 0.18 of the area subtended by our six cloud complexes in the solar circle, resulting in $\delta_a\sim 7.6\times 10^{-4}$. With this value, we get an upper bound for the pressure-regulated star formation rate per unit surface,
\begin{equation}
  \Sigma_{\rm SFR}(R) \approx 3 \times 10^{-3} \left(\frac {\Pism}{P_\odot }\right)T_{100}^{-1/2}~\text {M}_\odot~\text {kpc}^{-2} \text {yr}^{-1},
\end{equation}
  which, given the relationship between $\Pism$ and $\Sigma_{H_2}$ (Eq. \ref{eq:pressure_to_h2_is_cte}), can also be written in terms of the column density of molecular mass,
  \begin{equation}
   \Sigma_{\rm SFR}(R) \approx 3 \times 10^{-3} \left[\frac {\Sigma_{H_2}}{\Sigma_{H_2}(R_\odot)}\right]T_{100}^{-1/2}~\text {M}_\odot~\text {kpc}^{-2} \text {yr}^{-1}. 
\end{equation} 

\begin{figure}
    \centering
    \includegraphics[width=\columnwidth]{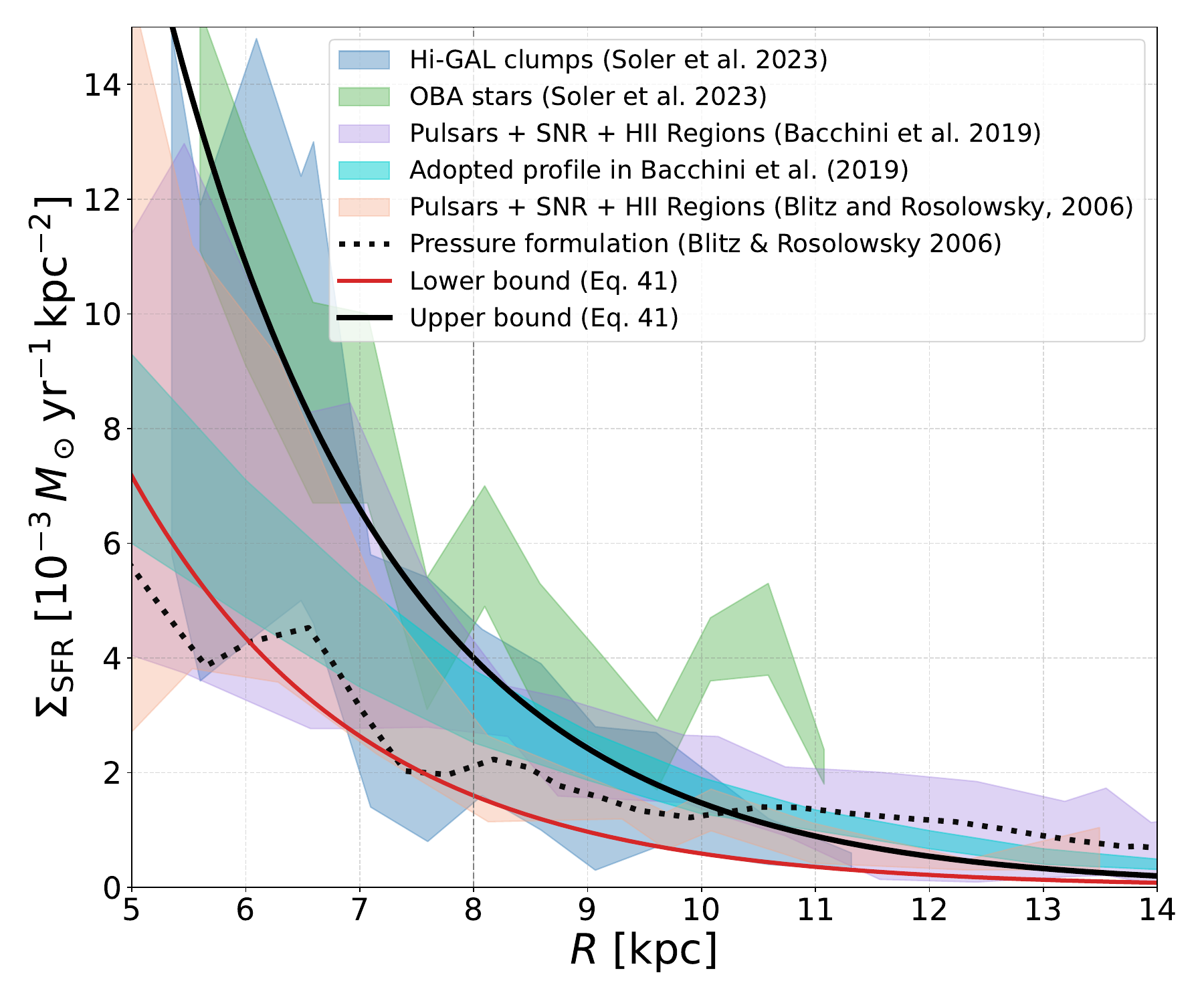} 
    \caption{Radial profiles of the star formation rate surface density, $\Sigma_\text{SFR}$, as a function of Galactocentric radius. The solid  curve represents the upper bound value in Eq. (\ref{eq:sfr_profile}) and the solid red line represents the lower bound. The blue and green shaded regions correspond to data from Hi-GAL clumps and OBA stars, respectively \citep[][see their Figure B.1]{Soler+2023}. The purple and beige regions represent estimates based on pulsars, supernova remnants, and HII regions, as reported by \citet{Bacchini+2019} and \citet{Blitz_Rosolowsky_2006}, respectively. The cyan region shows the adopted profile by Bacchini et al. (2019), while the dotted  line is the prescription proposed by \citet{Blitz_Rosolowsky_2006}.}
    \label{fig:sfr_density_profile}
\end{figure}

The values provided by these relationships and Eq. (\ref{eq:pressure_reg_sfr}) are within the limits of the observational radial distributions compiled by \citet{Misiriotis+2006,Blitz_Rosolowsky_2006,Lee_Miville_Murray16,Bacchini+2019,Palla+2020} and \citet{Soler+2023} for MW outside of 5 kpc, as shown in Figure \ref{fig:sfr_density_profile}. At the solar circle the observationally derived values are approximately $\Sigma_{\rm SFR}\sim~(1.5-5)~\times~10^{-3}$~M$_{\odot}$~kpc$^{-2}$~yr$^{-1}$ (note that the massive star tracers compiled by \citet{Soler+2023} give larger values for the whole Galactic disk but these depend on the assumed IMF). The typical temperatures of molecular clouds range between 10 and 50 K \citep{Snow_McCall2006}, using $T_{100}=0.5$ as a representative value of the average temperature at the onset of self-gravity, Eq. (39) gives $\Sigma_{\rm SFR\odot}\sim 4 \times 10^{-3}$ M$_{\odot}$ kpc$^{-2}$ yr$^{-1}$. Then, our upper and lower limits for the radial distribution of the surface density of the star formation rate in the MW can be written as,
\begin{equation} 
    \Sigma_{\rm SFR} \sim (1.5-4) \times 10^{-3}\ \text{exp} \left(\frac{R_{\odot}-R}{2 \  \text{kpc}}\right) M_{\odot} \ \text{kpc}^{-2} \text{yr}^{-1},
    \label{eq:sfr_profile}
\end{equation}
which, within the data scatter, follows the radial trends observed between 5 and 12 kpc in the MW. Note that the slope is within the limits derived by \citet{Lee_Miville_Murray16} for star formation in giant molecular cloud complexes, $\Sigma_{\rm SFR} \propto \exp{[-R/(1.7\pm 0.3 \ \text{kpc})]}$, and is also in line with the results of numerical simulations, using different virial parameters and CO conversion factors, discussed by \citet{Evans+22}. 

The resulting molecular gas depletion time for the MW disk is $t_{\rm dep}=\Sigma_{H_2}/\Sigma_{SFR}\sim (1-0.4)\times10^9$ yr, but this is a lower limit because the molecular gas is replenishing from the CNM reservoir at the rate given in Section (3.4). With respect to star formation inside spiral arms, the arm-interam pressure difference of a factor of 4 in the MW translates into a similar increase in the surface density of the star formation rate along the spiral arms. External galaxies show an average star formation contrast somewhat smaller, of a factor between 2 and 3 \citep{Querejeta+24}. 

A comparison of our rates with those obtained in nearby galaxies in the EDGE-CALIFA survey by \citet{Barrera-Ballesteros+21} and the ALMaQUEST survey by \citet{Ellison+24} shows a very good agreement for star-forming regions with pressures below log$(\Pism/k)\sim 5.5$. In Figure 5 we show the upper and lower bounds of the rates predicted by our model in Eq. (41), superimposed on the contours that surround 90, 80, and 50\% of the EDGE-CALIFA data obtained by \citet{Barrera-Ballesteros+21}. Their best fit to the data, which is almost linear, lies very close to our upper bound.

\begin{figure}
    \centering
    \includegraphics[width=\columnwidth]{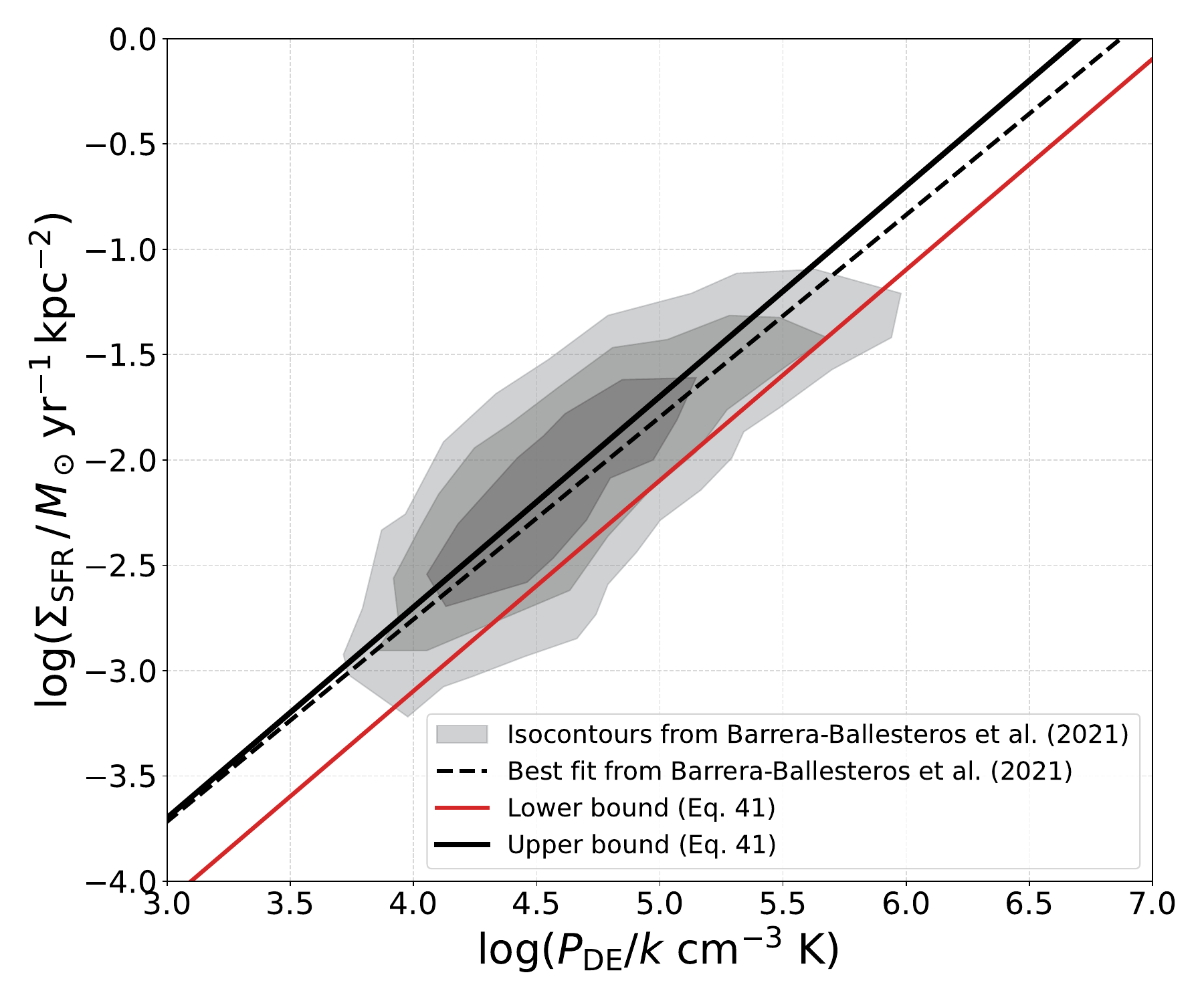} 
    \caption{Comparison of the rates predicted by our model with the results of the EDGE-CALIFA survey \citep[adapted from Fig. 9 in][graphed with the dynamic equilibrium pressure $P_{\rm DE}$, see Sect. 2.1]{Barrera-Ballesteros+21}. The shaded contours enclose the 90, 80, and 50\% data points of the sample and the dashed line represents their best fit.  The solid  curve is the upper bound value in Eq. (\ref{eq:sfr_profile}) and the solid red line is the lower bound.}
    \label{fig:sfr_EDGE-CALIFA}\
    \end{figure}
    
\subsection{\textbf{The Star Formation Efficiency along the disk of the MW}}
\label{sec:star_form_eff}

The transformation of a molecular mass into stars is a very inefficient process and proceeds at different speeds in different places. This has been a recurrent topic, discussed over the years in observational and theoretical studies, and the efficiencies derived from observations range from about 40 \% to a few percent \citep{Rydgren+1980,Mathieu1983,Rengarajan1984,Franco_Shore1984,Elmegreen_Clemens1985,Franco+1994,Young+1996,Williams+1997,Vazquez-Semadeni+2003,Zamora-Aviles+2019b, Wells+2022, Mattern+2024}. For isolated clouds, the star formation efficiency is defined as $\epsilon_{\rm sf}(t)=M_*(t)/M_0$, where $M_0=M_g(t)+M_*(t)$ is the initial mass of the star-forming cloud and $M_*(t)$ and $M_g(t)$ are the mass of recently formed stars and the remaining gas mass at any time $t$, respectively. Star formation is zero at the moment of cloud formation, and the efficiency with which the molecular mass is transformed into stars increases as time progresses. The maximum efficiency value is achieved at the end of the lifetime of the parental cloud, when it is finally evaporated by the combined action of expanding HII regions, stellar winds, and SNe. 

For the population of star-forming molecular clouds per unit area in a gaseous disk, the overall value of the final star formation efficiency in equilibrium is simply the ratio of the gas mass transformed into stars per unit time and unit area, $\Sigma_{SFR}$, to the rate of mass that is destroyed from the molecular cloud reservoir, $\dot{\Sigma}^{-}_{\rm mol}$ ,
\begin{equation}
\epsilon_{\rm sf}= \frac{\Sigma_{SFR}}{\dot{\Sigma}^{-}_{\rm mol}}\sim (3-8)\times 10^{-2}.
\end{equation}
Therefore, for pressure-regulated star formation in the MW, the efficiency is nearly constant along the disk and is similar to the ones derived by \citet{Franco+1994}, \citet{Federrath2016}, and \citet{Kim+21} from cloud evaporation by photoionization and the combined effects of turbulence, magnetic fields, and stellar feedback. Recent studies with young stellar object counts at nearby star-forming sites indicate that the efficiency decreases with the mass of the parental cloud,  and our derived value is appropriate for clouds with masses greater than $\sim 3\times 10^3 \ M_\odot$, where the efficiency tends to an asymptotic value of about 8 \% \citep [see Figure 10 in] [and references therein.] {Wells+2022}. 

With respect to the efficiency of star formation per free fall time \citep[see e.g.,][]{Krumholz+05,Krumholz+12}, our pressure-regulated scheme gives values that depend on the square root of the pressure,
\begin{equation}
\epsilon_{sf, ff}= \frac{\Sigma_{SFR}}{{\Sigma}_{H_2}} t_{ff}\sim (0.7-2)\times 10^{-2} \left(\frac{\Pism}{P_\odot }\right)^{1/2}T_{100}^{-1/2},
\end{equation}
which are compatible with those found by \citet{Kim+21} and with the values generally quoted in the literature \citep[see][and references therein]{Mattern+2024}. The observationally derived values of $\epsilon_{sf, ff}$ show large scatters, with mean values ranging from 0. 3 \% to 2. 6 \% \citep[see][]{Utomo+18,Mattern+2024,Leroy+2025, Meidt+2025}. The scatter is much smaller for the specific case of luminous and ultraluminous IR galaxies, where \citet{Wilson+2019} found larger but nearly constant mean values of 5\% and 7\%.  

\section{Discussion}

The interstellar pressure defines the properties of the gaseous components embedded in a galactic disk and is the substrate on which all physical and chemical interstellar processes occur, including those leading to molecular cloud formation and star formation. The velocity dispersion and temperature of the gas, its scale height, the mass exchange among the gas phases, and the destruction of star-forming clouds, on the other hand, are regulated by the energy input from young massive stars \citep[e.g.,][]{Field+1969,Cox_Smith1974,McKee_Ostriker1977,Cox1983,Franco_Shore1984,Parravano1988,Parravano1989,Elmegreen1989,Elmegreen_Parravano1994,Wolfire+2003,Blitz_Rosolowsky_2006,Ostriker+10,Colin+13,Elmegreen_Elmegreen2019,Grudic+21,Ostriker_Kim22}. Thus, interstellar pressure and star formation are intimately intertwined to keep the system stirred and in a self-regulated state. Our simplified model emphasizes the main physical processes that participate in their entanglement and provides the pressure-regulated rates that maintain equilibrium along the gaseous disk. 

The combined effects of differential rotation, supernova explosions, stellar radiation, and stellar winds drive the interstellar medium to form a network of gas flows across a wide range of scales. These flows easily develop internal interactions and can induce density fluctuations and hydrodynamical, thermal, and gravitational instabilities leading to the formation of molecular clouds and then to stars, but the conditions for each successive step towards star formation are different. In particular, for metallicities greater than 0.2 $Z_\odot$, the total column density required for the transition from a cold atomic cloud to a molecular cloud is approximately $2\Sigma_d \sim 10 \ \left(Z/Z_\odot\right)^{-1}~M_\odot$~pc$^{-2}$, which is equivalent to log$\ N_H\approx$ 21+log$\left(Z_\odot/Z\right)$. Aside from the solar neighborhood, this metallicity dependence is also appropriate to explain the observed gas column densities for the atomic-molecular transition regions in the Magellanic Clouds, where the average metallicities are $Z_{\rm LMC}~\sim~0.5~Z_\odot $ and $Z_{\rm SMC}~\sim~0.2~Z_\odot $ \citep[e.g.,][]{Roman-Duval+2022a}, and the FUV studies indicate log$\ (N_H)_{\rm LMC}\sim$ 21.3 and log$\ (N_H)_{\rm SMC}\sim$ 22 \citep[see][and references therein]{Shull+2021}. For the onset of internal gravitational instabilities, on the other hand, clouds must reach a column density of at least $\Sigma_{\rm sg} \sim 30 \ (\Pism/P_\odot)^{1/2}~M_\odot$~pc$^{-2}$. As is evident, the prerequisites for self-gravity and molecular cloud formation are different, but at the end, both are pressure-regulated. 

Our pressure-regulated scheme is in line and complements previous ideas on the role of interstellar pressure in regulating the properties of the gas and the star formation process\citep[e.g.,][]{Elmegreen1993, Elmegreen_Parravano1994, Wong_Blitz2002, Blitz_Rosolowsky04, Blitz_Rosolowsky_2006,Ostriker+10, Wilson+2019, Ostriker_Kim22}, and it can also be applied to a wide variety of different galactic environments as long as the metallicity is greater than 0.2 $Z_\odot$ and the system pressure is below log$(\Pism/k)\sim 5.5$. This is evident in the comparison with the results of the EDGE-CALIFA survey \citep{Barrera-Ballesteros+21} shown in Figure \ref{fig:sfr_EDGE-CALIFA}, which contains 4260 star-forming regions. These regions are distributed at different galactocentric radii along the disks of 96 nearby galaxies and have diverse local conditions with varied interstellar pressures, metallicities, gas velocity dispersions, radiation fields and cosmic ray fluxes. The upper limit of Equation \ref{eq:sfr_profile} lies very close to the best fit of the observational results, but both limits of the equation are within the range of observed values. In future work, we will address additional details of some of the observed properties. The pressure dependence derived in different studies varies somewhat, but is always close to unity for systems with low pressure. On the observational side, \citep{Blitz_Rosolowsky_2006} found a slope of 0.92 for MW and nearby spirals, but \citet{Fisher+2019} derived an exponent of 0.76 for the highly turbulent galaxies of the DYNAMO survey, and \citet{Molina+2020} obtained an exponent close to 0.8 for starburst galaxies, while \citet{Barrera-Ballesteros+21} for the EDGE-CALIFA survey and \citet{Ellison+24} for the extended ALMaQUEST sample got an almost linear exponent for systems with pressures below log$(P/k)\sim 5.5$, and a shallower relation for pressures above this value. On the theoretical side \citet{Ostriker+10} constructed a pressure-regulated model assuming that the injection of energy from $\Sigma_{SFR}$ should counterbalance the weight of the interstellar gas, and the results of the corresponding detailed numerical simulations by \citet{Kim+2011} and \citet{Ostriker_Kim22} yield exponents close to 1.2. \citet{Krumholz+2018} made a similar model but added gas inflows as an additional source of velocity dispersion and were able to get shallower relations at high rates of gas inflows that increase the interstellar pressures.

The model cannot be extended to the early phases of galaxy formation, but recent high-resolution hydrodynamic N-body simulations for galaxy formation are reaching parsec-size cell resolutions \citep{Bland-Hawthorn+2024,Bland-Hawthorn+2025}. This level of detail for possible MW progenitors, which is not necessarily achieved in cosmological hydrodynamical simulations, indicates that stellar feedback and gas pressure are coupled and also act in concert during the strong initial episodes of halo accretion and star formation. This coupling not only increases the velocity dispersion of newly formed stars but together with gravitational and shear instabilities in the nascent disk also drives bulk motions that shake the gas and stellar components, eventually leading to the formation of the stellar bar and the old stellar disk in the Milky Way progenitors. The possible importance of our work in this area lies in providing a robust phenomenological framework that can be easily implemented in such large simulated volumes, which can resolve the physics of molecular clouds in galaxy evolution simulations. 

For present-day conditions in the MW and spiral galaxies, dense clouds are typically formed through compressive events \citep{Hartmann+01, Vazquez-Semadeni+07}, and once a cloud becomes self-gravitating, it undergoes hierarchical collapse \citep[][and references therein]{Vazquez-Semadeni+19}. Consequently, the star formation rate must scale with the collapse rate \citep[see, e.g.,][]{Ballesteros-Paredes+24}, yielding the relation $\Sigma_{\rm SFR} \propto \Sigma_{\rm g} / t_{\rm sf}$, where $ t_{\rm sf}$ is the time it takes for a cloud to form stars, as several authors have discussed earlier \citep[e. g.][]{Larson1992,Wong_Blitz2002, Heiderman+2010,Krumholz+2012}. In our case, we restricted the participating gas mass at the onset of self-gravity to $\Sigma_{\rm sg}$ and the time scale to $t_{\rm ff}$, also at the onset of self-gravity. Note that our choice of time scale is in accordance with recent observational results in dense clouds. This fundamental dependence, assuming that self-gravitating clouds collapse on a free-fall timescale, allows us to account for the radial variation of the Milky Way's star formation rate. The absolute scaling of $\Sigma_{\rm SFR}$, particularly at the solar circle, is here constrained using the results from \citet{Bigiel+2008}, to derive a lower limit, and by the filling factor of star-forming clouds per unit area in the galaxy, which can be constrained by observations \citep[e.g.,][]{Miville-Deschenes+17, Roman-Duval+10, Evans+21, Spilker+21} to obtain the upper limit. Our upper limit values are almost in agreement with the best fit to the data of the EDGE-CALIFA survey derived by \citet{Barrera-Ballesteros+21}. 

In summary, metallicity defines the threshold condition for molecular clouds to form, and interstellar pressure sets the criterion for the onset of self-gravity. Once gravity takes over, the clouds collapse locally in about a free fall time, forming stars that will destroy their parental clouds, limiting the star formation process \citep[e.g., ][]{Franco+1994, Colin+13, Zamora-Aviles+19,Grudic+21, Kim+18, Ostriker+10,Ostriker_Kim22}. This scheme links the star formation rate with the gas pressure and is at the core of the global and hierarchical collapse scenario of molecular cloud evolution \citep{Vazquez-Semadeni+07,  Heitsch_Hartmann08, Ballesteros-Paredes+18, Vazquez-Semadeni+19}.

\section{Conclusions}

Here we have presented a simplified analytical model, based on the ideas first introduced by \citetalias{Franco_Cox86}, of the thresholds for molecular cloud formation and their transformation into self-gravitating entities, which are the actual sites of star formation. The results of this work are as follows. 

\begin{enumerate}
\item The onset of molecular cloud formation out of the diffuse CNM medium, as proposed by \citetalias{Franco_Cox86}, is mainly due to the opacity of the dust. For metallicities greater than $\sim$ 0.2~$Z_\odot$, the external HI-H$_2$ transition layer has a column mass density mainly of atomic gas $\Sigma_d\sim 5\ (Z_\odot /Z)\ M_\odot$ pc$^{-2}$.

\item We derive a modified version of the hydrogen molecule formation rate on dust grains that scales linearly with $\Pism$, $F\approx 3.16 \times 10^{-8}  \left( { \Pism}/{P_\odot}\right)\ T_{100}^{-1/2} \ \text{yr}^{-1}$. Given the threshold condition of the previous point, this implies that molecular cloud formation is controlled by pressure and metallicity. 

\item The total surface mass density threshold for molecular cloud formation and for star formation in the MW and nearby galaxies have a coincidental value of approximately $\sim 5-15 \ M_\odot$ pc$^{-2}$, but they refer to completely different entities. The one for the appearance of molecular clouds refers to twice the column density of the mainly atomic envelope protecting the molecular interior, $2\Sigma_d$. The one for the onset of star formation refers to the minimum molecular mass column density required to be self-gravitating. 
\item The formation rate of H$_2$ per unit area along the MW disk is $\dot{\Sigma}^{+}_{\rm mol}\sim 5\times10^{-2}\left(\Pism/{P_\odot}\right)T_{100}^{-1/2}M_\odot\text{kpc}^{-2}\text{yr}^{-1}$. Assuming equilibrium with the star formation rate, this is equal to the sum of the molecular mass transformed into stars plus the amount of molecular mass evaporated from the parental clouds due to the energy input from recently formed stars, per unit area and unit time, $\dot{\Sigma}^{-}_{\rm mol}$. 
\item The star formation process inside $\sim$11~kpc follows a two-step path. First, a molecular cloud is formed out of cold neutral gas, and when it contracts or gathers more gaseous mass from the external medium, it becomes self-gravitating and then is able to form stars.
\item   The star formation rate per unit surface on the MW disk, $\Sigma_{\rm SFR} \sim (1.6-4) \times 10^{-3}\ (\Pism/P_\odot )\  M_{\odot} \ \text{kpc}^{-2} \text{yr}^{-1}$, is pressure regulated and scales equally well and linearly with the surface mass density of molecular hydrogen. This result is in very good agreement with the EDGE-CALIFA and ALMaQUEST surveys for log$(\Pism /k)\leq 5.5$. The corresponding star formation efficiency is $\epsilon_{\rm sf}\sim(3 - 8)\times 10^{-2}$ and the efficiency per free fall time is $\epsilon_{\rm sf, ff}\sim(0.7-2)\times 10^{-2}(\Pism/P_\odot )^{1/2}T_{100}^{-1/2}$.

\end{enumerate}

\section*{Acknowledgements}

The authors thank Stan Kurtz, Sebastián Sánchez, Steve Shore and an anonymous referee for very useful comments and suggestions that helped us to improve the present work. JF acknowledges many lively and informative discussions on these topics and over the years with Don Cox, Gilberto Gómez, Tom Hartquist, Gerhard Hensler, Jong Soo Kim, Francesca Matteucci, Ron Reynolds, Sergei Silich, and Enrique Vazquez-Semadeni.  ARP acknowledges financial support from DGAPA-PAPIIT IN106924.
JBP acknowledges financial support from the UNAM-DGAPA-PAPIIT IG101723 grant.

\section*{Data Availability}

No new data was generated during the preparation of this paper.



\bibliographystyle{mnras}
\bibliography{Bibliography, MisReferencias} 
\bsp	
\label{lastpage}

\end{document}